# *On Dynamic Price Formation in the Course of Capital Reallocation Driven by Differential Rates of Profit: Strict Conservation of Value Supports Karl Marx's Theory*

Norbert Ankri[a] and Païkan Marcaggi[b]

[a] Independent researcher, Marseille, France; formerly INSERM, Aix-Marseille University, Marseille, France

[b] Muséum National d'Histoire Naturelle, CNRS UMR7196, INSERM U1154, Paris, France

**Corresponding author:** Norbert Ankri, norbert.ankri@univ-amu.fr

**Abstract**

We develop a three-sector model in which Marx's aggregate equalities (total price equals total value; total profit equals total surplus value) restrict a sequence of fully realized states during capital reallocation and technical diffusion. Anchoring one sector's profit rate to the contemporaneous average selects a trajectory without modeling pricing intentions. Before innovation, capital inflow lowers the receiving sector's price and excess profitability along calibrated stable paths, without a demand function. Stability is predictable from value and technical data; near a stable equalization, the receiver's profit rate falls even where its price rises. A labor-saving, capital-intensifying innovation creates temporary extra surplus value and revalues committed capital. At constant real wages the final uniform rate rises, as Okishio predicts; when rising wages keep exploitation constant, it falls.

## 1. Introduction

The transformation of values into prices of production remains a central controversy in Marxian economics. This paper studies one precise proposition: what follows dynamically if Marx's two aggregate equalities—total price equals total value and total profit equals total surplus value—are imposed as a strict conservation law throughout competitive adjustment. We consider the conservation of value across the full circuit of capital, from physical production to market sale, on the explicit assumption that no crisis of realization occurs. The validity of this law is a matter of debate. On the one hand, Marx explicitly formulates the two aggregate equalities, and they remain central to some modern reformulations (e.g., Freeman 2020; Moseley 2016). On the other hand, Maurice Lagueux (1984) argues that they only make sense through a principle of value conservation, which cannot hold in Marx's theory as a whole: "a conservation principle seeks to establish that nothing of what is said to be conserved is either lost or created," whereas "value is created in social labor and lost" through unproductive consumption or devaluation (94; our translation). He grants, however, that the principle "could apply without restriction" to the transformation of surplus value into profit, which concerns value already created (96). That is our domain: living labor creates value in production; circulation conserves it. His decisive objection, after Bortkiewicz (1907), is that at a uniform rate of profit the two equalities hold together only in exceptional cases (Lagueux 1984, 99). Here they hold at every step, and convergent paths end in such a case (§5.2). Nor did Poincaré's questioning of the substantialist reading of conservation principles, which Lagueux invokes (91), abolish them: energy conservation became the statement that "there is something which remains constant" (Poincaré 1902, chap. 8; our translation). In this non-substantialist sense, we treat Marx's two aggregate equalities axiomatically. The purpose is not to settle every interpretation of Marx's value theory, but to test the internal coherence and dynamic consequences of a value-conserving price system.

The model also deliberately abstracts from fixed capital, credit, inventories and unequal turnover times. A failure under these favorable conditions would reveal an inconsistency;

success establishes coherence under stated assumptions, not a description of an actual economy.

We use a three-sector circulating-capital model with bread and meat as wage goods and iron as a means of production. Sraffa's physical system supplies the uniform equilibrium profit rate for a given technique and real wage, while labor values determine surplus value and the contemporaneous value-based average rate along the adjustment path. At uniform-profit equilibrium the two rates coincide. Off equilibrium, while sectoral profit rates are still dispersed but the technique is unchanged, labor values remain well defined and continue to determine a contemporaneous average rate that closes the price system.

For a given capital allocation, the two conservation equations constrain three transformation coefficients and leave one degree of freedom. A third equation selects one point of this admissible set at each allocation, and hence a trajectory. We place one sector's profit rate at the contemporaneous average and compare two closures that converge from the initial allocation. This mathematical selection does not assume that firms intend to earn the average rate. Different pricing strategies or demand conditions could select other realized configurations, but, under the conservation hypothesis, these configurations must satisfy the same aggregate constraints. Closure therefore selects among compatible states rather than merely fixing a numéraire (Shaikh 2016).

Ankri (forthcoming) shows that, once both aggregate equalities are imposed, capital allocation is constrained rather than arbitrary in the two-sector case. Ankri and Marcaggi (2026) extend the argument to three sectors under simple reproduction. Here competition through capital mobility selects a path within the admissible set, an innovation diffuses through coexisting techniques, committed capital is revalued, and the two aggregate equalities are checked at every recorded state.

This dynamic setting produces three main results. First, before innovation, capital inflow expands the receiving sector's output and, under both convergent closures, lowers its unit price and excess profitability without an imposed excess-supply price equation. We show that this response reflects the gap between surplus value produced and the profit a uniform rate would require, and derive an ex ante condition for the stability of the adjustment. Near equalization, we prove, for any viable three-sector technique with only basic commodities (Supplementary Material S1.9), that convergent paths exist, that this condition is equivalent to a fall in the receiver's price relative to the donor's, and that a stable approach lowers its profit rate even where its price rises. Second, a cost-reducing innovation generates temporary extra surplus value for the new technique while devaluing capital committed to the innovating sector. Third, distribution governs the long-run profit-rate response. With a constant real wage, the uniform rate rises, as Okishio's (1961) theorem predicts. With a real-wage path leaving aggregate exploitation unchanged, organic composition rises and the uniform rate falls. The latter result complements general analyses showing that viable capital-using, labor-saving technical change can lower the equilibrium rate when exploitation is held constant (Basu and Orellana 2023) and analyses under alternative distributive assumptions (Chen 2023; Basu 2026). Our distinctive contribution is not another existence proof, but the value-conserving transitional path, including heterogeneous techniques, capital revaluation and reproducible iteration-level accounting.

The Python code reproduces all numerical results and Figures 1–4, reports conservation residuals, and permits alternative parameters and anchors. Section 2 defines the model and its closure; Section 3 specifies capital mobility and innovation diffusion; Section 4 reports the simulations; Section 5 discusses their scope and limitations; Section 6 concludes.

## 2. Simulation Model and Parameters

To enable comparison with previous work (Basu 2021), we use a three-sector economy producing bread (sector 1), iron (sector 2), and meat (sector 3). The technical structure is defined by:

Input–output matrix A: each element $a_{ij}$ represents the physical quantity of commodity *i* required to produce one unit of commodity j:

$$A = \begin{pmatrix} a_{11} & a_{12} & a_{13} \\ a_{21} & a_{22} & a_{23} \\ a_{31} & a_{32} & a_{33} \end{pmatrix} = [a_{ij}]_{i,j=1,2,3}$$

Direct labor row vector ℓ: $\ell_j$ denotes the number of labor hours directly required to produce one unit of commodity j:

$$\ell = (\ell_1, \ell_2, \ell_3)$$

Real-wage (subsistence) bundle v, a column vector: $v_i$ denotes the quantity of commodity i purchased by workers per hour of labor paid as wages. In the baseline case, workers consume only bread and meat; superscript T indicates transposition:

$$\mathbf{v} = (v_1, 0, v_3)^T$$

**The augmented socio-technical matrix Ã:**

$$\tilde{A} = A + v\ell$$

$$\tilde{A} = [\tilde{a}_{ij}]_{i,j=1,2,3},\ \tilde{a}_{ij} = a_{ij} + v_i\ell_j$$

vℓ is the product of the real-wage vector and the direct-labor vector. It incorporates wage-good requirements. For example, in bread production, direct labor $\ell_1$ entails wage-good requirements $(v_1\ell_1, 0, v_3\ell_1)^T$ in the baseline specification. The theoretical uniform profit rate r* follows from the dominant eigenvalue of Ã (Perron–Frobenius theorem).

The model contains no fixed capital: all capital advanced is circulating capital, consumed within the period.

*2.1 Pre-Innovation Parameters*

$$A=\begin{pmatrix} 0.4 & 0.45 & 0.4 \\ 0.08 & 0.30 & 0.266667 \\ 0.1 & 0.15 & 0.1 \end{pmatrix} \quad \ell = (0.06, 0.035, 0.04) \qquad \mathbf{v} = (2, 0, 1)^T$$

The labor value embodied in each unit of commodity *i*, denoted $\Lambda_i$, is determined by solving the system:

$$\Lambda_j = \Sigma_i \Lambda_i a_{ij} + \ell_j, \text{ for } j = 1, 2, 3$$

Provided the Hawkins–Simon conditions are satisfied, the matrix (I−A) is invertible and, in matrix notation:

$\Lambda = \ell(I - A)^{-1}$, where I is the identity matrix

This equation yields the unique pre-innovation value vector $\Lambda = (\Lambda_1, \Lambda_2, \Lambda_3)$.

For our numerical example:

$\mathbf{\Lambda} = (0.15221582, 0.18352604, 0.16647407)$

The unit of account is socially necessary labor time; one unit of value represents one hour of labor.

The pre-innovation augmented socio-technical matrix is:

$$\tilde{A} = \begin{pmatrix} 0.52 & 0.52 & 0.48 \\ 0.08 & 0.30 & 0.266667 \\ 0.16 & 0.185 & 0.14 \end{pmatrix}$$

r* = 0.2033

*2.2 Post-Innovation Parameters (Sector 2)*

The innovation in the iron-producing sector 2 increases productivity by reducing direct labor input ($\ell'_2$=0.03 instead of $\ell_2$= 0.035) at the cost of a slightly higher iron requirement, representing capital intensification ($a'_{22}$=0.305 instead of $a_{22}$=0.3, changed values in bold; primes denote post-innovation values). The real wage v remains unchanged.

$$A' = \begin{pmatrix} 0.4 & 0.45 & 0.4 \\ 0.08 & 0.305 & 0.266667 \\ 0.1 & 0.15 & 0.1 \end{pmatrix} \qquad \ell' = (0.06, \mathbf{0.030}, 0.04) \qquad \mathbf{v} = (2, 0, 1)^T$$

The new vector Λ′ and new augmented matrix $\tilde{A}'$ are calculated using the formula $\Lambda' = \ell'(I-A')^{-1}$ and $\tilde{A}' = A' + v\,\ell'$:

Λ’ = (0.15074737, 0.17608461, 0.16361656)

$$\tilde{A}' = \begin{pmatrix} 0.52 & 0.51 & 0.48 \\ 0.08 & 0.305 & 0.266667 \\ 0.16 & 0.18 & 0.14 \end{pmatrix}$$

r*′ = 0.2055

*2.3 Sectoral Decomposition of Capital*

Let $K_j$ denote the capital committed to sector j and $k_j$ the capital advanced per unit of commodity j, both in value, and let $Q_j$ be the number of units produced, so that $K_j = Q_j k_j$, with $k_j = \Lambda_1 a_{1j} + \Lambda_2 a_{2j} + \Lambda_3 a_{3j} + \ell_j \Lambda v$.

As in the commodity-by-commodity decomposition of Sraffa’s (1960) system, we break $k_j$ =

$b_j + f_j + m_j$ down into its bread, iron ('f' for Fe, from the Latin *ferrum*) and meat components, which include the wage goods: $b_j = \Lambda_1 \tilde{a}_{1j} = \Lambda_1(a_{1j} + v_1 \ell_j)$, $f_j = \Lambda_2 \tilde{a}_{2j} = \Lambda_2 a_{2j}$ and $m_j = \Lambda_3 \tilde{a}_{3j} = \Lambda_3(a_{3j} + v_3 \ell_j)$.

The surplus value per unit is $s_j = \Lambda_j - k_j$. The rate of exploitation is $e = (1 - \Lambda v)/(\Lambda v)$ and, for sector j, $e_j = s_j/[(\Lambda v)\ell_j]$. Given a common wage bundle v, in any state with a homogeneous technique $e_j = e$ for every sector: from the definitions above, $s_j = \Lambda_j - k_j = \ell_j(1 - \Lambda v)$, so $e_j$ reduces to $(1 - \Lambda v)/(\Lambda v)$ for every j — a consequence of the model's structure in such states, not an added assumption. During innovation diffusion, when two techniques coexist in Sector 2 (§3.2), the socially recognized value Λ reflects a mixture of the old and new techniques rather than either one exactly, while each sub-branch produces at its own unit cost; the measured $e_j$ therefore departs transiently from uniformity, and the economy-wide rate below becomes the operative measure. This departure is an accounting effect of the recognition lag, not a difference in the labor actually performed.

$C_j$ and $V_j$ denote the constant and variable capital committed to sector j, in value: $C_j = Q_j(\Lambda_1 a_{1j} + \Lambda_2 a_{2j} + \Lambda_3 a_{3j})$ and $V_j = Q_j \ell_j \Lambda v$.

Economy-wide exploitation is $\bar{e}$ = S_total / V_total = $\Sigma_j V_j e_j / \Sigma_j V_j$, the variable-capital-weighted mean of the sectoral exploitation rates. This coincides with e whenever $e_j$ is uniform across sectors and becomes the operative definition once innovation diffusion (§3–§4) introduces sectoral heterogeneity in $e_j$.

Sectoral organic composition is $OC_j = C_j/V_j = \Sigma_i \Lambda_i a_{ij} / [(\Lambda v)\ell_j]$. Their economy-wide counterparts are C_total = $\Sigma_j C_j$, V_total = $\Sigma_j V_j$ and OC_total = C_total/V_total. Before innovation, $(OC_1, OC_2, OC_3)$ = (3.264, 9.012, 6.714); for the post-innovation technique at the unchanged wage basket, (3.252, 10.470, 6.644). Although sector 2 becomes more capital-intensive (higher organic composition), the capital advanced per unit falls ($k_2$: 0.16501 → 0.16004) because of the large fall in variable capital per unit.

This sectoral decomposition of capital allows the Marxian conservation constraints to be expressed in a *disaggregated* form (by commodity type rather than in constant and variable capital) linking value categories directly to the physical input–output structure.

*2.4 Model Closure: Sectoral Profit-Rate Anchoring Method*

The average profit rate $\bar{r}$ at any time t is calculated prior to price formation as the ratio of total surplus value to total committed capital, both in labor-value terms:

$\bar{r} = \bar{r}(t) = \Sigma s_j Q_j(t) / \Sigma k_j Q_j(t)$

This value-based measure is computed from the current state before prices are formed and serves as a predetermined reference for the price system (see below). $\bar{r}$ equals the theoretical rate r* (calculated from the dominant eigenvalue of $\tilde{A}$) only once profit rates $r_1$, $r_2$ and $r_3$ have equalized.

Each iteration represents one discrete step of competitive adjustment: the price system is recalculated, profit-rate differentials are observed, and a small quantum of capital is reallocated from the least profitable sector to the most profitable one. The sequence is historical rather than temporal: iterations are ordered and irreversible, each state depending on the one before, but they correspond to no fixed interval of clock time. The number of iterations required for profit rates to equalize therefore measures the length of an adjustment path, not its duration.

The adjustment is quasi-static: each iteration records a state in which the produced quantities are assumed to have been sold and productive inputs and wage goods can be replaced. The three-equation system determines its selected prices and enforces Marx’s two aggregate equalities. This realization equilibrium must be distinguished from uniform-profit equilibrium. Sectoral profit rates can still differ after the commodities have sold, and their differences drive the subsequent reallocation of capital. Price fluctuations, shortages and unsold output during the intervening market adjustment are outside the recorded sequence.

The dynamic, iterative anchoring procedure is applied as follows:

1. At any given iteration t, the economy-wide average profit rate $\bar{r}$ is computed from the current value aggregates using the above equation.

2. The profit rate of a chosen sector (bread or meat) is anchored to this contemporaneous average $\bar{r}$.

3. This constraint, combined with the two aggregate equalities, yields a unique set of transformation coefficients, prices and profit rates for iteration *t*.

4. Capital then migrates in response to profit rate differentials, updating sectoral sizes and value aggregates for the next iteration *t*+1.

Let $x_i = p_i/\Lambda_i$ be the transformation coefficient of sector i, the ratio of its unit price to its unit value. These coefficients are determined, along with the prices of production, as a result of three constraints (the two aggregate equalities and the anchoring of one sector's profit rate to $\bar{r}$) detailed as follows.

**Total value–price equality:**

$$W_1 x_1 + W_2 x_2 + W_3 x_3 = W_1 + W_2 + W_3 \quad \textbf{(1)}$$

where $W_j = Q_j \Lambda_j$ is the value of sector j's output

**Surplus-value equality (S = Π),** with capital advanced (wage goods included) resolved by commodity type:

$$(W_1 - B) + (W_2 - F) + (W_3 - M) = S = \Pi = (W_1 - B)\, x_1 + (W_2 - F)\, x_2 + (W_3 - M)\, x_3 \qquad \textbf{(2)}$$

where $B = \Sigma j\,(Q_j b_j)$, $F = \Sigma j\,(Q_j f_j)$, $M = \Sigma j\,(Q_j m_j)$, $S = \Sigma j\,(Q_j s_j)$

$S$ and Π are respectively total surplus value and total profit.

The profit of branch j is: $\Pi_j = W_j\, x_j - Q_j\,(b_j\, x_1 + f_j\, x_2 + m_j\, x_3)$

where $b_j$, $f_j$, $m_j$ are the per-unit values of the bread, iron and meat advanced in branch *j*, inclusive of wage goods, as defined in §2.3. The corresponding branch profit rate is:

$$r_j = \Pi_j / Q_j\,(b_j\, x_1 + f_j\, x_2 + m_j\, x_3)$$

The $r_j$ are equalized to $\bar{r}$ only at uniform-profit equilibrium; off equilibrium they differ and drive capital reallocation. Because prices redistribute value, $\Pi_j$ generally differs from the branch's surplus value $Q_j s_j$, while $\Sigma_j \Pi_j = S = \Pi$.

These equalities are imposed as a conservation law on gross aggregates, not as a net-product normalization (see Supplementary Material S4.1).

Throughout the dynamic simulation, we verify that the physical reproduction constraints hold at each iteration, i.e., $(W_1 - B) \geq 0$, $(W_2 - F) \geq 0$, $(W_3 - M) \geq 0$. This ensures the internal consistency of our reproduction scheme under the assumption of no initial stocks. In a more general setting, temporary deficits could be covered by inventories from prior periods; their sustained absence here confirms that the adjustment path we model is self-sustaining. The constraint is an inequality, not an equality: a strictly positive net product remains in each commodity, so simple reproduction is not assumed. Note also that, given equation (1), equation (2) is equivalent to $Bx_1 + Fx_2 + Mx_3 = B + F + M$, that is, to the conservation in prices of the value of total capital advanced. Marx's two aggregate equalities therefore entail a third aggregate invariant, which our disaggregated formulation makes explicit.

**Price equation of the anchored sector (sector a):**

$W_a x_a = (1 + \bar{r})(B_a x_1 + F_a x_2 + M_a x_3)$ **(3)**

where $B_a = Q_a b_a$, $F_a = Q_a f_a$, and $M_a = Q_a m_a$

Equation (3), written for the anchored sector (here, sector 1 or 3), closes the system and selects a unique price vector for each recorded allocation. Other selections within the admissible set are possible; the anchor lets us study a determinate sequence of realized states and its stability.

Real economies are less orderly: production cycles differ, sales are uncertain, and firms may target higher margins or cut prices to gain market share. We set these processes aside; our object is the set of realized states compatible with the conservation hypothesis, and whether adjustment through them is viable.

## 3. Dynamic Process: From Production Equilibrium to Profit-Rate Equalization

The model's starting state, defined by A, $\ell$, v and initial capital allocation $K_{init} = (K_1, K_2, K_3)_{init}$ (each $K_j$ measuring capital committed to sector *j*), represents a reproducible production state: the gross output of each sector covers the entire demand addressed to it as a productive input and as a wage good, so that all capital advanced is materially replaced. The residual net product, the material bearer of surplus value, is assumed to be fully realized, whether through capitalist consumption, accumulation, or exchange at equivalent value with a market outside the circuit considered; the model does not track its destination, since our object is the formation of prices of production, not a schema of accumulation. This commodity-reproduction condition is entirely compatible with a disequilibrium in profit rates across sectors. It is precisely this inter-sectoral inequality in profitability that triggers the competitive movement of capital. The following phase formalizes this adjustment process, in which capital flows from sectors with below-average profit rates to those with above-average rates, driving the system toward a uniform rate of profit.

*3.1 Phase 0 – Profit-Rate Equalization through Capital Reallocation under Constant Technology (Iterations 2–200)*

Total invested capital $K_T = K_1 + K_2 + K_3 = 1000$ with $K_j = Q_j\, k_j$. Starting from the initial condition $K_{init}$ = (560, 260, 180), capital migrates from lower- to higher-profit sectors according to a discrete adjustment rule:

$$\Delta K_h = \alpha \cdot (r_h - r_l)/r_l$$

where l and h are the least and most profitable sectors at each iteration. This rule conserves the total value of capital invested ($K_T$), as the outflow from sector l equals the inflow to sector h ($\Delta K_l = -\Delta K_h$). Coefficient $\alpha = 10$ is measured in the same units as the capital variables $K_j$. This parameter, chosen by way of illustration, primarily affects the trajectory and the convergence time, but its value does not alter the qualitative results of the transition or the final equilibrium, except in the event of instability. Transfers are switched off once $\max_j |r_j - \bar{r}| / \bar{r} < 10^{-10}$. Adjustment rules of this type, in which capital flows respond to observed profit-rate differentials rather than presupposing an already established uniform rate, follow the line of classical disequilibrium analysis developed by Duménil and Lévy (1987), where gravitation toward a uniform rate is the outcome of a competitive process and its stability is established only under specified conditions. For the parameterization used here, this stability can also be characterized analytically. If h and l denote the more and the less profitable of the two non-anchored sectors, z the capital of h and $g = r_h - r_l$, then local stability around z*, the capital allocation at which profit rates are equalized (r*), requires $-2r^*/\alpha < g' < 0$, where g′ is the derivative of g with respect to z at z*; the derivation and verification for both convergent anchors are given in Supplementary Material S1.3–S1.4.

The corresponding limiting profit-rate-equalized capital allocations K* are:

- **When sector 1 (bread) is anchored:**

K* = (560, 251.3904152, 188.6095848), $\bar{r}$ = 0.2032511 = r*

With the bread anchor, the relative dispersion falls below $10^{-6}$ at iteration 131 and stands at $2.6 \times 10^{-9}$ at iteration 200, the end of Phase 0. Because the anchored profit rate is set equal to the contemporaneous average $\bar{r}$, which is a positive weighted average of the sectoral rates, sector 1 cannot be the least or the most profitable unless all rates are equal; under our min-to-max transfer rule it therefore never participates in capital migration. Transferring only between the extreme branches is a simplification that does not affect our conclusions.

- **When sector 3 (meat) is anchored:**

K* = (561.9018018, 258.0981982, 180), same $\bar{r}$ = 0.2032511 = r*

Convergence is far quicker here: the dispersion falls below $10^{-6}$ at iteration 18 and below $10^{-10}$ at iteration 30, where transfers cease. Here it is $K_3$ that stays fixed, for the same structural reason. Anchoring the innovating Sector 2 instead makes Phase 0 diverge before any innovation occurs: the transformation coefficients drift and $r_1$ turns negative. The divergence is a failure of the branch-level convergence criterion (see Supplementary Material S1.3): capital becomes cumulatively self-attracting. The failure depends on the allocation as well as on the closure: from allocations with a smaller iron share, on the appropriate side of its singular line, the iron anchor converges (Supplementary Material S1.9). A divergent path cannot be a sequence of realized states. From $K_{init}$, rates and prices become aberrant: by iteration 10, $r_1 = -0.43$ and meat sells at 2.47 times its value. From the equalized allocation, the shrinking meat sector soon cannot supply the meat that the expanding branches require, and its net product turns negative (Supplementary Material S1.3). Forces the model does not describe, such as demand, pricing behavior or the shortages themselves, must then restore an admissible selection, such as that of a convergent anchor.

The resulting equilibrium, where Marx's equalities are strictly satisfied, is documented with a full sectoral breakdown in Supplementary Tables S1 and S2 for iteration 199 (bread anchor).

*3.2 Phase 1 – Price Calibration and Innovation Diffusion*

All value magnitudes are functions of the iteration t. We suppress the argument where no ambiguity arises and retain it only where two different time indices occur in the same expression: in the law of motion of θ, in the interpolation rule for Λ, in the revaluation of $K_2$ and in the wage schedule, $v(t) = v(t-1) + v_3_add$.

*Initial price calibration*

For each determined K*, the initial transformation coefficients are reported in Supplementary Material S2.1 (they differ only marginally between the two anchors).

The progressive diffusion process is modeled as follows:

*Innovation diffusion (iterations 201–301)*

At iteration 201, a cost-reducing innovation begins in sector 2 (iron production). The innovation modifies both the technical coefficient matrix from A to A' and the direct labor vector from ℓ to ℓ', raising $a_{22}$ from 0.300 to 0.305 (slightly more constant capital per unit) and lowering $\ell_2$ from 0.035 to 0.030 (less labor time per unit).

The innovating firms are assumed to satisfy Okishio's cost-reduction criterion: the new technique must lower the unit production cost when both techniques are evaluated at the price vector prevailing before the innovation is generalized. In the present model, these prices are given by $p_i = x_i\Lambda_i$. Okishio (1961) showed that, at a given real wage, a technique that is cost-reducing in this sense cannot lower the uniform rate of profit. The theorem covers capital-using, labor-saving (Marx-biased) techniques such as the present one.

If the innovation is introduced in sector 2, the criterion is written as:

$$[pA' + (pv)\ell']_2 < [pA + (pv)\ell]_2$$

which holds here: at pre-innovation prices the unit cost of iron falls by 0.78 percent, so the investment is rational at existing prices. All three unit values fall ($\Lambda' < \Lambda$), consistent with Roemer's (1981) analysis of viable technical change. The subscript (2) denotes the second (iron) component.

*Dual technology structure and progressive diffusion*

Unlike the simpler approach in Phase 0, we now model sector 2 as split into two coexisting technique-vintages, each operating a fraction of the sector's capital, throughout the diffusion period: a fraction $1 - \theta$ of its capital (technique α) operates the old technology $(A, \ell)$, and the complementary fraction θ (technique β) the new one $(A', \ell')$.

This heterogeneity captures the realistic scenario where early adopters coexist with laggards, creating temporary competitive advantages that drive further diffusion. The share of Sector 2 capital operated with the new technology follows a linear diffusion path described by the parameter $\theta \in [0,1]$, evolving from 0 (iteration 201) to 1 (iteration 301). We emphasize that θ is a share of the sector's capital, not a proportion of firms: the model contains no firm count, and the language of adopters and laggards is used only as motivation. Nor is θ a share of output (Supplementary Material S3.1).

This yields a heterogeneous sector where both techniques coexist, each with distinct unit costs:

$k_{2\alpha} = b_{2\alpha} + f_{2\alpha} + m_{2\alpha}$ (old technology)

$k_{2\beta} = b_{2\beta} + f_{2\beta} + m_{2\beta}$ (new technology)

Since the point of the innovation is to reduce the capital invested per unit of output, we have $k_{2\beta} < k_{2\alpha}$. In the real world, this difference is what drives diffusion of the innovation.

*Socially necessary labor time and value dynamics*

A distinction is drawn between the physical rate of adoption (linear increase of θ) and the redetermination of socially necessary labor time, which Λ_calc approximates. This adjustment is set through a one-step interpolation:

For t ≥ 201, we first compute an interpolated technical structure:

$A_eff(\theta) = (1-\theta)A + \theta A'$

$\ell_eff(\theta) = (1-\theta)\,\ell + \theta\ell'$

from which we calculate $\Lambda_calc(t) = \ell_eff \times [I - A_eff]^{-1}$.

However, the operative labor values Λ(t) used in value calculations are obtained by interpolation:

For t ≤ 200, θ(t) = 0 and Λ(t) = Λ

For 201 ≤ t ≤ 301, θ(t) = (t – 201)/100

and $\Lambda(t) = \Lambda_calc(t-1) + \theta(t) \times [\Lambda_calc(t) - \Lambda_calc(t-1)]$

For $t > 301$, $\theta(t) = 1$ and $\Lambda(t) = \Lambda'$

Using the direct rule $\Lambda(t) = \Lambda_calc(t)$ changes the average profit rate by at most $3.1 \times 10^{-6}$ and final total capital by less than $10^{-3}$. The reported findings therefore do not depend on a particular recognition lag.

*Surplus value and capital allocation*

Despite technological heterogeneity, both fractions of Sector 2 capital produce iron of the same socially recognized unit value, $\Lambda_2$.

However, their costs differ, yielding differential surplus values:

$s_{2\alpha} = \Lambda_2 - k_{2\alpha}$

$s_{2\beta} = \Lambda_2 - k_{2\beta}$

Since $k_{2\beta} < k_{2\alpha}$, innovating firms capture extra surplus value: $s_{2\beta} > s_{2\alpha}$. This extra surplus constitutes a temporary technological rent that persists only during the diffusion phase.

The mechanics of the diffusion (the division of Sector 2 capital between the two techniques, the aggregation of its exploitation rate, the revaluation of $K_2$ as unit costs change, and the optional real-wage schedule) are set out in Supplementary Material S3.

*Non-conservation of total capital*

Unlike Phase 0, where inter-sectoral capital flows preserve KT, Phase 1 exhibits a net release of capital. This release, which occurs despite temporary super-profits in Sector 2 and is quantified in §4.1, mirrors Marx's analysis of the release and devaluation of capital (Marx [1894] 1981, ch. 6). Two channels are at work. First, the innovation is capital-saving per unit ($k_{2\beta} < k_{2\alpha}$), so the aggregate unit capital $k_2 = K_2/Q_2$ falls as $\theta$ rises, since $Q_2 = [1-\theta]K_2/k_{2\alpha} + \theta K_2/k_{2\beta}$; because committed capital rather than output is carried forward here, this channel raises output. Only Sector 2's capital is revalued; elsewhere falling unit costs raise output instead. The release comes from the second channel: $\Lambda_2$ falls from 0.18353 to 0.17608, devaluing Sector 2's capital under both techniques. Inter-sectoral capital flows are driven by profit rate differentials, redistributing capital between sectors, but these flows merely reallocate a shrinking total; they do not create the shrinkage itself. The reduction in $K_T$ thus combines technical revaluation with the exogenous wage path.

This release of capital does not result from a contraction of physical output: over the same interval, output rises, with iron increasing from 1523.5 to 1550.1 units (+1.75 percent) and wage goods by different proportions (+1.53 percent bread, +1.03 percent meat). What falls is not output itself, but the value of committed capital required to obtain it. The subsequent use of the released capital lies outside the scope of the present paper.

*Profit rate dynamics and value conservation*

During diffusion (e.g., at iteration 250), profit rates diverge across sectors. The innovating sector 2 experiences temporarily elevated profitability due to reduced costs, inducing capital inflows from other sectors. Despite these price and profit rate fluctuations, the model maintains the two aggregate equalities:

$\Sigma_i W_i = \Sigma_i W_i x_i$ and $\Sigma_i S_i = \Sigma_i \Pi_i$

These equalities are imposed at each iteration.

*3.3 Numerical Implementation and Independent Cross-Validation*

All results reported below were produced by the Python reference implementation value_conservation_simulation.py accompanying this paper: 450 recorded states per scenario (iteration labels 2–451), including 199 Phase-0 states, the diffusion window 201–301, and 150 post-diffusion states, with a capital-migration tolerance of $10^{-10}$. Across the four scenarios and 1,800 recorded states, the maximum conservation residuals are: total price minus total value, exactly zero; total profit minus total surplus value, $2.3 \times 10^{-13}$; anchored rate minus $\bar{r}$, $5.0 \times 10^{-16}$; minimum reproduction surplus, 4.92 (this smallest net product remains comfortably positive).

The model was implemented twice independently — in LabVIEW and in Python, with different linear solvers — and the two agree on every equilibrium value to within $3.1 \times 10^{-6}$ in capital allocation and eight significant figures in profit rates. Figures are drawn from the Python reference. This dual implementation is part of the reproducibility claim: the conclusions do not depend on any particular numerical library.

As $\theta \to 1$ and diffusion completes (iteration 301), sector 2 has fully adopted the new technology and extra surplus value disappears; profit rates then re-equalize. Labor values $\Lambda_j$ stabilize at their post-innovation levels, reflecting the now-generalized productivity increase. The temporary technological rent vanishes as the innovation becomes universal, a result consistent with Marx's analysis of relative surplus value (Marx [1867] 1976, ch. 12) under competitive conditions.

## 4. Results

### *4.1 Profit-Rate and Capital-Allocation Dynamics (Figures 1 and 2)*

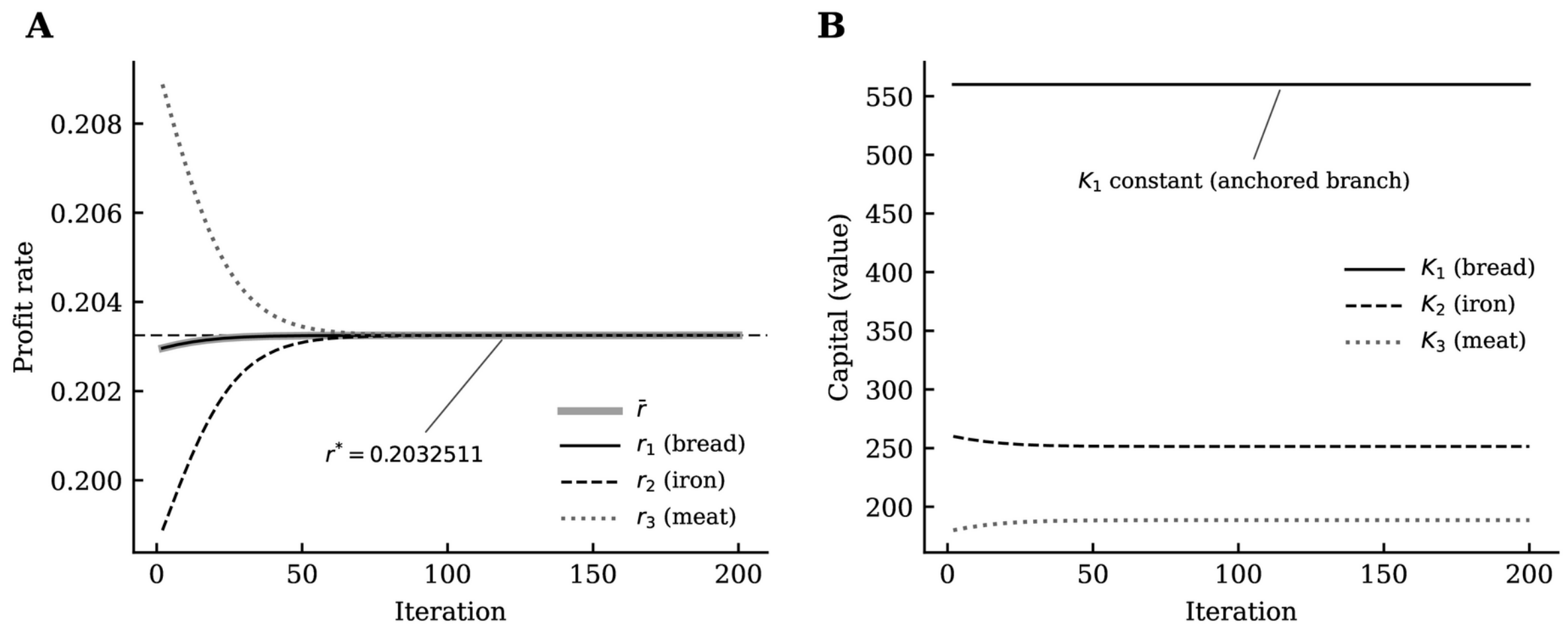

**Figure 1. Profit-rate equalization and capital reallocation during Phase 0 under the bread anchor.** Panel A shows sectoral rates converging to r* = 0.2032511; the anchored rate $r_1$ coincides with the average. Panel B shows capital moving from iron to meat while $K_1$ remains fixed. Lines: bread, solid; iron, dashed; meat, dotted gray; average, thick pale gray. Source: Authors' simulations.

Figure 1 illustrates profit and capital allocation dynamics during Phase 0. Panel A shows how sectoral profit rates converge toward the theoretical profit rate r* as a consequence of the movement of capital from the branch with the lowest profit rate (branch 2) to the branch with the highest (branch 3). In this example, sector 1 was chosen as the anchor (i.e., its profit rate is constrained to equal the economy-wide average $\bar{r}(t)$ at every iteration). Panel B shows how capital allocation gradually adjusts as profit rates converge.

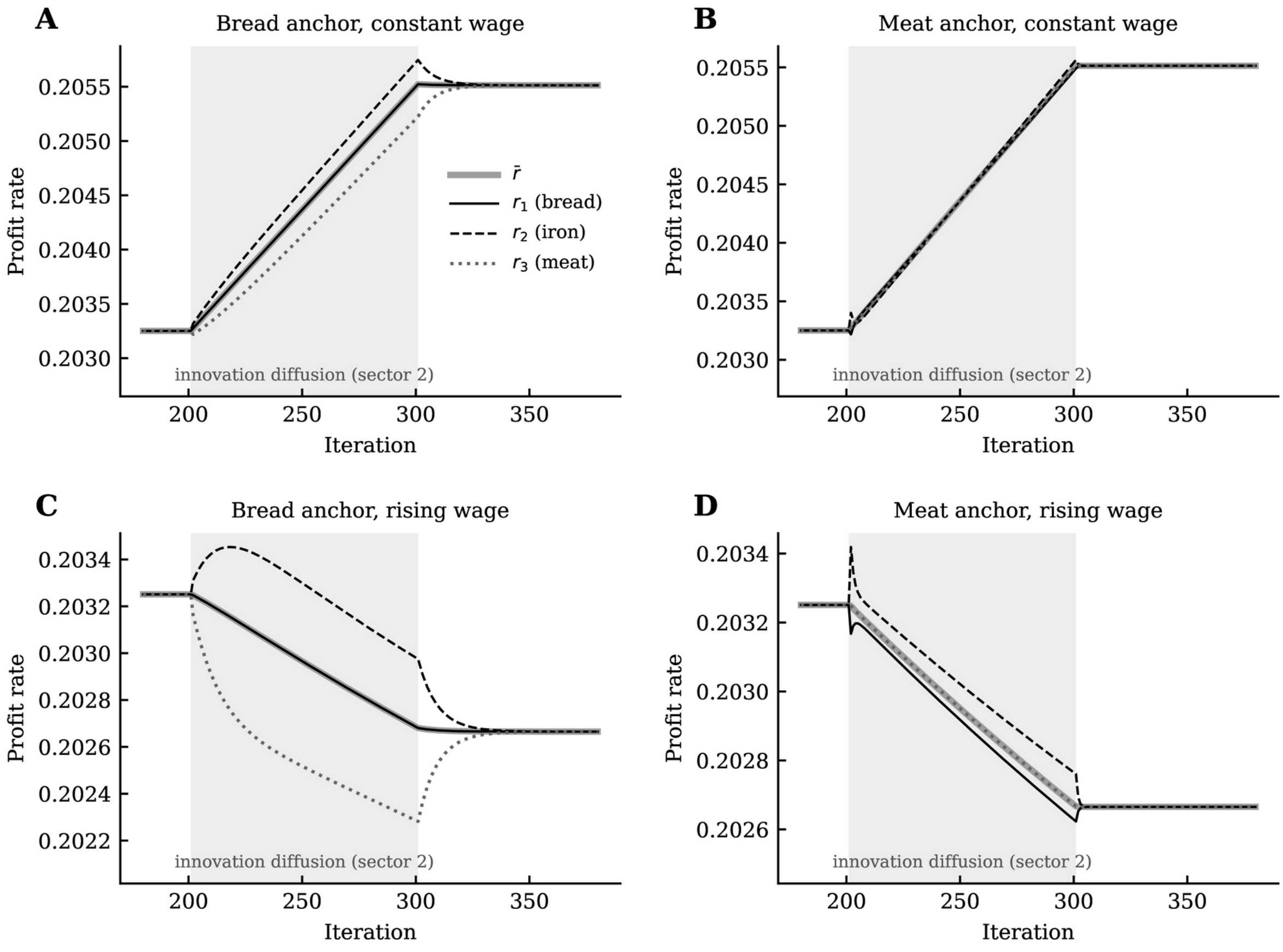

**Figure 2. Profit-rate dynamics during innovation diffusion.** Panels A–B report constant-real-wage scenarios under bread and meat anchors; Panels C–D report the corresponding rising-wage scenarios. The shaded interval is iterations 201–301. The average rate rises to 0.20551 at constant wages and falls to 0.20267 with the specified wage increase. The anchored rate coincides with the average. Source: Authors' simulations.

Figure 2 illustrates the behavior of sectoral profit rates during Phase 1 when different sectors are chosen as the anchor. Panels A and B show the effect of innovation diffusion while the real wage remains constant. Panels C and D show the effect while the real wage increases progressively as a result of increased productivity.

Panel A shows the case where Sector 1 (bread) is anchored: $r_1$ remains identical to $\bar{r}$ throughout the simulation, while Sectors 2 and 3 exhibit transient deviations during the innovation diffusion phase (iterations 201–301). The innovation reduces both labor time ($\ell_2$: 0.035 → 0.03) and capital per unit of iron output ($k_{2\beta} < k_{2\alpha}$). Although Sector 2's elevated profitability during early diffusion attracts capital flows from Sector 3, this influx is insufficient to offset the progressive devaluation induced by the innovation itself. Consequently, $K_2$ declines overall from 251.39 (pre-innovation equilibrium) to 248.08 (post-innovation), while $K_3$ falls from 188.61 to 186.50. Total capital declines from $K_T = 1000$ (pre-innovation) to approximately 994.58 (post-innovation). Across the four scenarios, final total capital ranges from 994.46 to 994.85 depending on the anchor and wage regime. Sector 2's profit rate stays above the average throughout diffusion, reflecting the extra surplus value captured by firms deploying the new technology; once the innovation has generalized, migration re-equalizes all rates. The average profit rate $\bar{r}$ rises from approximately 0.20325 to 0.20551 between pre- and post-innovation equilibria, consistent with Okishio's theorem being satisfied at constant wages.

Panel B presents the symmetric case where Sector 3 (meat) is anchored, again with a constant real wage. Here, $r_3$ tracks $\bar{r}$ by construction, yet the sectoral patterns closely resemble those in Panel A. The economy-wide average rates of Panels A and B differ by at most $9.1 \times 10^{-6}$ during diffusion and coincide at both equilibria: the anchor affects relative prices transiently, not the average rate.

Panels C and D reproduce the same anchoring configurations (bread and meat, respectively) but now incorporate progressive real wage increases during the innovation phase ($v_3$_add = 0.000354 per iteration; cumulative $\Delta v_3 = 0.0354$). Wage increments begin one iteration after the innovation decision, which is taken at unchanged prices and unchanged real wages, so Okishio's criterion is evaluated before any wage response. The contrast with Panels A and B is stark: although the innovation satisfies Okishio's criterion throughout, the rising wage trajectory produces a declining average profit rate $\bar{r}(t)$ throughout the diffusion period, falling from approximately 0.20325 to 0.20267. Cost-reducing innovation at prevailing prices thus need not raise profitability once wages rise, a case outside Okishio's fixed-real-wage hypothesis.

The close alignment of Panels C and D confirms that this time path reflects the wage dynamics, not the choice of anchor, with stable profit-rate levels before and after diffusion.

*4.2 Exploitation Rate Dynamics and Technological Rent (Figure 3)*

The exploitation rate profiles of Figure 3 are invariant to the choice of profit-rate anchor: exactly so for every sectoral rate, and to within $8.2 \times 10^{-6}$ in relative terms for the aggregate $\bar{e}$ (§5.1), a result that parallels the profit-rate invariance demonstrated in Figure 2. This invariance is not coincidental: it follows directly from our value-conservation framework, where the two aggregate equalities ensure that value-based categories ($e_j$, $\Lambda_j$) are determined independently of profit rate anchoring. Consequently, the class-conflict dynamics of exploitation (measured in value terms) are intrinsic to the production and wage conditions, not artifacts of the transformation procedure.

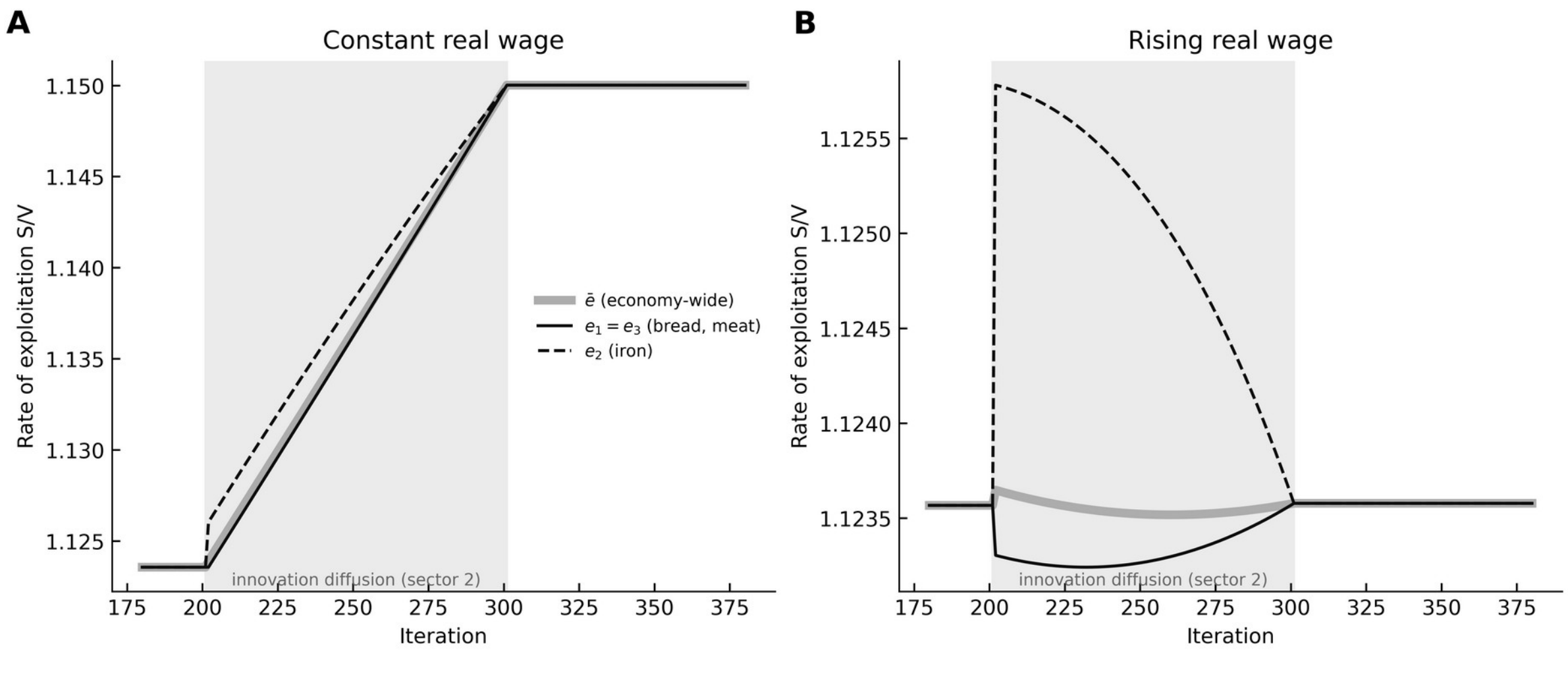

**Figure 3. Sectoral and aggregate exploitation rates under constant and rising real wages.** Bread-anchor results are shown for iterations 180–380. In Panel A, the constant wage permits exploitation to rise after the productivity gain. In Panel B, wage growth keeps aggregate exploitation approximately constant while the innovating sector receives a temporary rent. The shaded interval is iterations 201–301. Source: Authors' simulations.

Figure 3 decomposes aggregate exploitation dynamics to reveal the mechanism of technological rent appropriation and its differential sectoral incidence. The spike in Sector 2's exploitation rate ($e_2$) at first positive adoption (t = 202) reflects the capture of extra surplus value by early adopters: while both old and new iron-production techniques produce output valued at identical socially necessary labor time ($\Lambda_2$), the innovation's lower unit costs ($k_{2\beta} < k_{2\alpha}$) generate higher surplus per unit ($s_{2\beta} > s_{2\alpha}$). This rent is intrinsically transient: both panels show $e_2$ reconverging toward $\bar{e}$ as diffusion progresses ($\theta \rightarrow 1$). However, the trajectory of reconvergence diverges fundamentally between wage regimes, with non-trivial implications for non-innovating sectors that static analyses obscure.

Panel A (Constant real wage): The pre-innovation equilibrium has $e \approx 1.124$. During diffusion, Sector 2's rate jumps to $e_2 = 1.12605$ at the onset and then rises monotonically to 1.15002, which is the new common equilibrium level rather than a transient maximum (a true peak occurs only when real wages rise; Panel B). The critical feature, however, is that the exploitation rates of sectors 1 and 3 also rise from 1.124 to the same final equilibrium, reflecting system-wide productivity spillovers: cheaper wage goods lower the value of labor power $\Lambda v$, raising $e = (1 - \Lambda v)/(\Lambda v)$ and $e_1$ and $e_3$ despite the absence of technical change in branches 1 and 3. Sector 2's elevated path captures the redistributional moment (early-adopter advantage), but this advantage dissipates as the technology becomes universal, leaving only the system-wide productivity gain. The aggregate rate $\bar{e}$ rises smoothly by 2.4 percent under constant wages even though technical change is confined to Sector 2.

Panel B (Rising real wage): Sector 2's small but distinct spike to $e_2 \approx 1.1258$ (about 0.2 percent above baseline) reflects a temporary technological rent before wage adjustments materialize. Yet as diffusion proceeds concurrently with rising wages (v_add), $e_2$ subsequently declines. Sectors 1 and 3 experience a slight temporary decline in exploitation

during mid-diffusion, falling below their pre-innovation baseline. A full sectoral breakdown is given in Supplementary Tables S3 and S4, for iteration 251 ($\theta = 0.5$, bread anchor with rising wages). This counterintuitive dip emerges from the temporal asymmetry between wage increases and productivity spillovers: variable capital ($V_1$, $V_3$) rises immediately with wages, while the benefit of cheaper iron inputs accrues gradually through the value system's progressive adjustment mechanism. The measured exploitation rate for non-innovating sectors transiently falls, despite economy-wide technical progress. By the end of the simulation, rising wages fully absorb productivity gains, leaving $\bar{e}$ essentially unchanged, indicating a near-complete wage-led neutralization of surplus value extraction.

Theoretical implications: The economy-wide relationship $\bar{r} = \bar{e}/(OC_total + 1)$ holds exactly throughout the dynamics. In Panel A, rising $\bar{e}$ combines with capital devaluation to raise $\bar{r}$ (0.20325 → 0.20551), satisfying Okishio's theorem. Panel B's near-constancy of $\bar{e}$ means wage growth offsets the productivity gain: both C_total and V_total fall, but V_total falls proportionally more, so OC_total rises and $\bar{r}$ declines (0.20325 → 0.20267). Thus, a cost-reducing innovation is accompanied by falling profitability when workers capture productivity gains. The temporary dip in $e_1$ and $e_3$ reveals a crucial dynamic absent from static frameworks: the asymmetry runs between the innovating branch and the others and originates in diffusion, a rising wage shifting only the level common to all branches. The wage path is imposed uniformly and should not be interpreted as a model of branch-level bargaining (§5.5).

## 5. Discussion: Aggregate Invariants, Price Formation, and Sectoral Dynamics

### *5.1 Invariance, Conservation, and Macroeconomic Validity*

Value categories are largely invariant to the profit-rate anchor. Sectoral exploitation rates are identical across anchors, while the economy-wide average profit-rate paths are nearly invariant (Figure 2). Total surplus value's invariance is an analytic consequence of $S = \Pi$ at every state. Anchor choice can nevertheless alter sectoral transients. For instance, under meat anchoring, Sector 2's profit rate briefly falls below the average after the initial innovation spike (Figure 2B), in contrast to its steady increase above the average under bread anchoring. The aggregate relation $\bar{r} = \bar{e}/(OC_total + 1)$ remains an exact identity at every iteration because total surplus value equals total profit and total value equals total price.

Between bread and meat anchoring, $e_1$, $e_3$, $e_2\alpha$, $e_2\beta$, $e_2$ and $\Lambda(t)$ agree to machine precision, and the equilibrium average rate matches $r^*$ to eight digits. Along the transition the two anchors no longer coincide, since total surplus value itself depends on the allocation once the coefficients $s_j/k_j$ vary with $\theta$. During diffusion $\bar{e}$ differs by at most $9.2 \times 10^{-6}$ and $\bar{r}(t)$ by $1.5 \times 10^{-5}$; the largest deviation of the whole path occurs early in Phase 0, where $\bar{r}$ differs by $2.1 \times 10^{-4}$ at iteration 6 before the paths converge.

Geometrically, this invariance has a simple reading. With total committed capital fixed, the admissible allocations form a simplex — the triangle of non-negative $K = (K_1, K_2, K_3)$ summing to $K_T$. On this triangle, total surplus value is the scalar product $S = \Sigma_j(s_j/k_j)K_j$, since $Q_j = K_j/k_j$. Because there is no fixed capital (§5.5), the ratios $s_j/k_j$ are fixed by the technique and the wage basket, so S is linear in K: the loci of equal surplus value are parallel straight lines, all orthogonal to the projection of the vector $(s_j/k_j)$ onto the triangle. Profit-rate equalization confines the allocation to one of these lines, the one on which $\bar{r} = r^*$, which passes through the proportions of the standard system (§5.2; Supplementary Material S1.8); moving along it leaves S, and hence $\bar{r} = S/K_T$, unchanged — the two anchors merely select different points on it.

This bears directly on Steedman's (1977) claim that value magnitudes are superfluous once the physical system is given, and on Freeman's (2020, n. 18) counterclaim that price is redundant. At uniform-profit equilibrium, the equivalence is exact. Writing p for the row vector of unit prices and $\tilde{A} = A + v\ell$, uniform profitability gives $p = (1 + \bar{r})p\tilde{A}$, so that $1/(1 + \bar{r})$ is the dominant eigenvalue of $\tilde{A}$ and $\bar{r}$ is the Sraffian uniform rate. From the value side, Marx's equalities imply $S = \Pi$ and, jointly, $Bx_1 + Fx_2 + Mx_3 = B + F + M$; therefore $\bar{r} = S/(B + F + M) = \Pi/(Bx_1 + Fx_2 + Mx_3)$. The physical and value constructions thus necessarily yield the same uniform rate. Nor is the system overdetermined: at Sraffa's prices only the scale is free, so the two equalities can hold together only if $S = r^*K_T$, a condition on the allocation of capital rather than on prices (§5.2).

A distinction becomes sharper off equilibrium. Marx's two aggregate equalities impose a numéraire-independent condition on relative prices themselves: a single normalization can enforce one aggregate equality but not, in general, both — a constraint the static physical system alone does not supply. Sraffa does not predict a particular off-equilibrium price path; the derivation is given in Supplementary Material S4.4.

In Phase 0 the contemporaneous average rate converges to the Sraffian equilibrium rate as capital is reallocated; during diffusion, where two techniques coexist and no single augmented matrix describes the economy, $\bar{r}$ departs from the interpolated-technique rate by at most $2.1 \times 10^{-5}$ (numerical detail in Supplementary Material S4.4). Value accounting therefore both defines a contemporaneous average rate and, with the adjustment rules, restricts the historical path of relative prices and profits while preserving the aggregates redistributed by transformation.

As Duménil (1982) notes, in a Morishima-type system the composition of the real-wage basket, not only its value, affects profitability; this is why the real wage enters here as a physical vector rather than a scalar share (numerical illustration in Supplementary Material S4.5).

*5.2 Price Formation and Surplus-Value Redistribution (Figure 4)*

The model addresses the compatibility of prices and quantities in realized states, not the faster transactions through which supply and demand meet.

At fixed technique, $Q_i = K_i/k_i$, so capital inflow expands the receiving sector's output. Yet $Q_i$ cancels from the profit-rate ratio (§2.4): at unchanged relative prices, output, aggregate cost and profit would grow proportionally and the rate would remain unchanged. Reallocation instead changes the commodity weights in the two aggregate equations and the surplus value produced, so the next realized state requires different transformation coefficients. Under both convergent closures, the receiving sector's output rises while its unit price and excess profitability fall, as in classical gravitation analyses (Duménil and Lévy 1987; Bellino and Serrano 2018).

Conservation applies within each state, between its value and price accounts; it does not keep total surplus value constant between successive states. Even with total capital fixed, reallocation changes the living labor employed and therefore S: the adjustment reorganizes both the creation and the appropriation of surplus value.

What uniform profitability requires appears when the sectoral rates $r_j$ are averaged in two ways. Weighted by the shares $\varphi_j$ of capital advanced at current prices, they average $\bar{r} = S/K_T$ (§2.4). Weighted by the shares $\varphi^*_j$ of the standard system of $\tilde{A}$ at those prices, they average $r^*$ at any price vector, because that system, taken as a whole, reproduces its means of production and wage goods in the physical ratio $1 + r^*$ (Sraffa 1960). Hence:

$$\Sigma_j (\varphi^*_j - \varphi_j)\, r_j = r^* - \bar{r} \qquad (4)$$

Identity (4) uses the two aggregate equalities alone, not the anchor. Uniform rates therefore require $S = r^*K_T$ whatever rule selects prices: the allocation must lie on the line of §5.1, and prices must be Sraffa's. In Phase 0, where all sectors share the rate of exploitation e, $\rho_j = s_j/k_j = e/(1 + OC_j)$ and $\bar{r} = e/(1 + OC_total)$: the social capital must have the organic composition of the standard system, $OC^* = e/r^* - 1 = 4.528$ (against $OC_total = 4.536$ at $K_{init}$), at which Marx's general rate equals Sraffa's (§5.1). Off the line, let $\Gamma = r^*K_T - S$. With sector a anchored at $\bar{r}$, the receiving (h) and donating (l) sectors have equal and opposite excess profits relative to remuneration at the current average; the receiver's is $(r_h - \bar{r})C_h^p = \Gamma/(\sigma_h - \sigma_l)$, where $\sigma_j$ compares the sector's weight in the standard system with its actual weight (Supplementary Material S1.8). This excess differs from the gap between a branch's appropriated profit and its own surplus value; redistribution of surplus value continues at uniform rates.

At $K_{init}$, capital sets 383.61 hours of living labor in motion, against 384.15 consistent with uniform rates, giving $\Gamma = (1 - \Lambda v)(384.15 - 383.61) = 0.29$. Under the bread anchor, moving capital from iron ($OC_2 = 9.01$) to meat ($OC_3 = 6.71$) lowers the social organic composition and raises total surplus value. The narrowing of Γ outweighs the change in proportions: the local derivatives give an 11.6 percent reduction in Γ and an 8.4 percent reduction in the receiver's premium $r_h - \bar{r}$ per unit moved (Supplementary Table S5). Between $K_{init}$ and equalization, meat output grows by 4.783 percent but its profit by only 2.039 percent; its profit per unit falls enough to outweigh a small rise in unit cost, lowering its price by 0.388 percent (Figure 4A–B).

The response depends on the selected trajectory. Under the iron anchor, the transfer from bread ($OC_1 = 3.26$) to meat raises the social organic composition, widens Γ by 41 percent, raises meat's price and destabilizes the adjustment (Figure 4C–D). Under the migration rule of §3.1, for the anchor closure and any viable three-sector technique in which every commodity is basic, the uniform-profit allocation is locally stable, for small enough α, if and only if $(\rho_a - r^*)(\kappa_a - \kappa^*_a) > 0$, where $\kappa_a$ and $\kappa^*_a$ are the anchored sector's actual and standard capital shares. In Phase 0 this reads $(OC^* - OC_a)(\kappa_a - \kappa^*_a) > 0$: an anchored sector of lower organic composition than the standard system must be over-represented in the social capital, one of higher composition under-represented. The condition holds at $K_{init}$ for bread and meat, but not for iron. Generically, at every equalized allocation except the standard one, some anchor satisfies the condition and another violates it (Supplementary Material S1.9.3). Conservation thus guarantees that stable paths exist without selecting one; forces outside the model, such as demand, can select it. Convergent paths exist for every such technique whose value rates $\rho_j$ are not all equal: for small enough $\alpha$, each equalized allocation satisfying the condition attracts a whole interval of initial allocations. Near equalization, the receiver's price relative to the donor's falls if and only if the condition holds. A stable approach lowers the receiver's profit rate from either side, even where its absolute price rises, and lowers that price from at least one side. Supplementary Material S1.9 gives detailed proofs of these results; the simulations illustrate them. The proofs are local, but the receiver's rate also fell in all 13,365 stabilizing transfers computed for random techniques and allocations. The direction of movement is that of classical gravitation analyses, but the explanation differs: no excess-supply price rule is involved. Near equalization, the aggregate equalities and the anchor fix the change in profit rates, and the change in prices follows. These rate and price results extend to n sectors when all other sectors are anchored at the average rate.

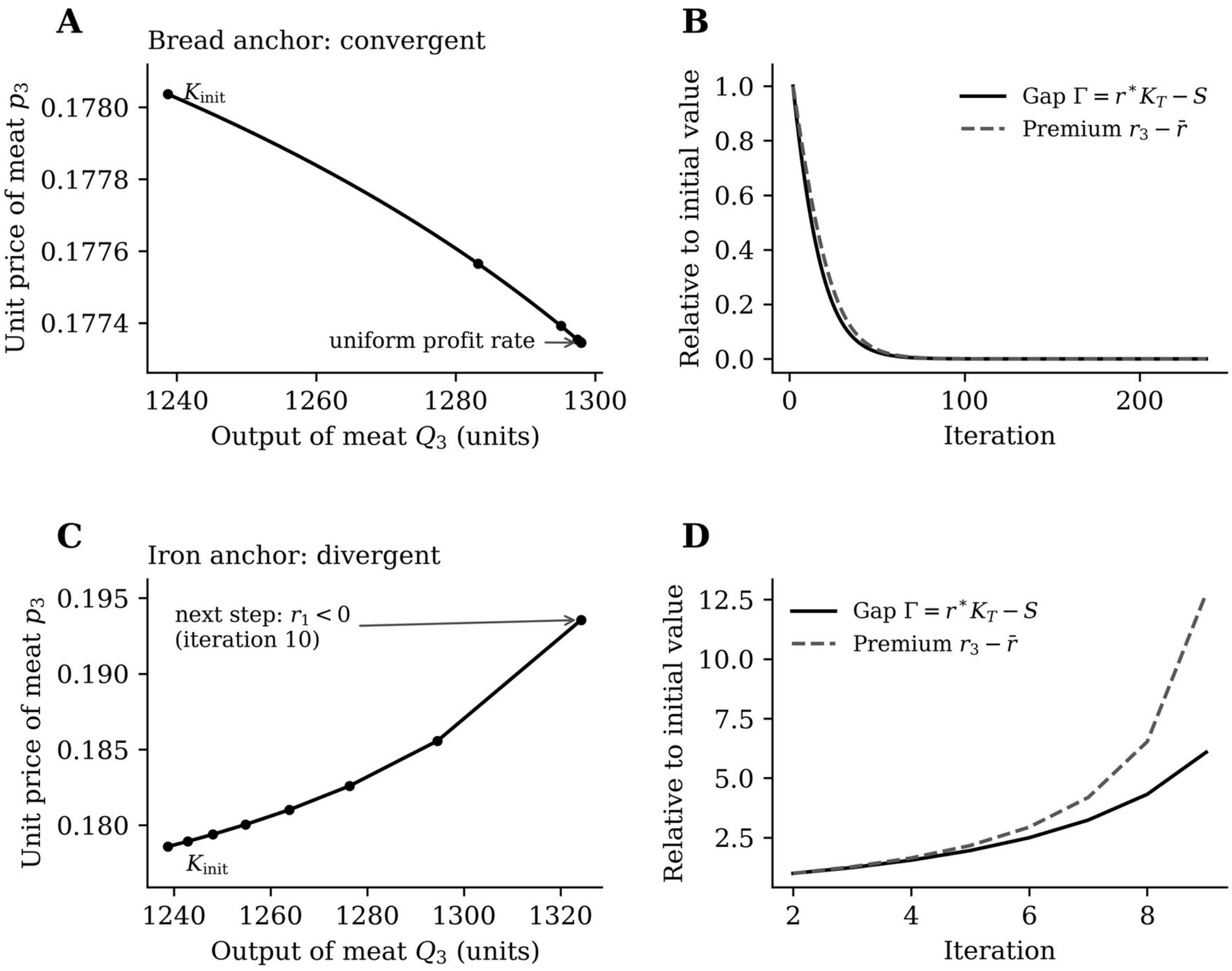

**Figure 4. Price response of the receiving sector and the surplus-value gap.** Phase 0 from $K_{init}$ under the migration rule of §3.1; meat receives capital under both anchors. Panels A and C: unit price of meat $p_3$ against its output $Q_3$. Panels B and D: gap $\Gamma = r^*K_T - S$ between the profit a uniform rate would require and total surplus value, and premium $r_3 - \bar{r}$ of meat, each relative to its initial value. Under the bread anchor (A–B) the gap closes, the premium vanishes and the price falls as output rises. Under the iron anchor (C–D) the gap widens, the price rises with output and the adjustment diverges: $r_1$ turns negative at iteration 10. Each state is assumed fully realized, so demand is taken to have absorbed each of these price-quantity pairs; the locus is not, however, a demand curve. Source: Authors' simulations.

This construction distinguishes two timescales without supplying a behavioral model of the faster one: demand, preferences or pricing strategies may move realized prices within the admissible set before capital reallocates. Husson (2017) objects to Laure van Bambeke (2013) that "the time of capital reallocation is not, in any case, that of price formation" (our translation). Market realization can precede profit-rate equalization: supply and demand have met in a recorded state, yet unequal profitability can induce the next capital transfer. The construction specifies compatible states and the stability of the selected sequence, rather than a demand curve or a transaction-level law of price adjustment (Supplementary Material S4.6).

*5.3 Technological Rent and Asymmetric Sectoral Effects*

Figure 3 also shows innovation's temporary, asymmetric impact on non-innovating sectors, which comparative statics cannot display. Under constant wages, $e_1$ and $e_3$ rise throughout diffusion but stay below $e_2$ until the innovation has generalized; under rising wages they dip below their pre-innovation level before reconverging, while Sector 2's temporary rent ($e_2 \approx 1.1258$) erodes as diffusion proceeds.

Supplementary Tables S1 and S2 make this transfer legible. At the pre-innovation equilibrium Sector 1 produces 147.57 of surplus value but appropriates only 112.40, while Sector 2 produces 28.21 and appropriates 51.97, some 84 percent more than it created. The redistribution is therefore not an effect of innovation but the ordinary operation of the transformation, which allocates surplus value in proportion to capital advanced rather than to labor employed; innovation merely widens a transfer that is already structural.

*5.4 Profit Rate Dynamics and the Limits of Okishio's Framework*

Technological superiority at prevailing prices does not guarantee rising profitability, as the divergence between wage regimes shows. Under constant wages (Figure 2, Panels A–B), the cost-reducing innovation raises $\bar{r}$, confirming Okishio's theorem within its restrictive framework; under rising wages (Panels C–D), exploitation stays essentially constant and the rise of OC_total from 4.528 to 4.544 (§4.2) drives $\bar{r}$ downward, as Marx argued: the wage dynamic dominates the devaluation of capital and the reduced labor time per unit. Temporal single-system critiques of the theorem rely on historical-cost valuation (Kliman 2007). With circulating capital advanced anew at each state and slow diffusion, our simultaneous valuation is its quasi-static limit: the rate rises at a constant real wage, and its fall requires the wage path. General sufficient conditions for such an outcome have been derived elsewhere (Basu and Orellana 2023; Chen 2023; Basu 2026). Okishio's criterion also excludes higher-cost, more capital-intensive techniques chosen to enlarge the mass of profit.

### *5.5 Scope and Limitations*

**Realization.** The reproduction inequality ensures that gross output covers productive and wage-good requirements with a positive net product. Full sale of this remainder through capitalist consumption, accumulation or exchange outside the circuit is a separate assumption. Accumulation alone cannot absorb the whole remainder for the reported allocations: its maximum balanced-growth rate depends on the composition of the net product. A model of effective demand and inventories would be needed to describe realization failures and the market adjustment between recorded states. The present sequence instead identifies states compatible with conservation, conditional on realization. Our valuation differs from the Temporal Single System Interpretation (TSSI; Kliman 2007) chiefly by supplying the equilibrium reference it rejects and a determinate price path during capital reallocation; Supplementary Sections S4.1–S4.2 discuss realization and accumulation ceilings.

Two distinct failures of realization can be distinguished. If unsold output is destroyed, the labor expended is never validated as value; only the sold quantity, and the correspondingly smaller surplus value, is recorded — a straightforward narrowing of the accounting base. If output sells at a depreciated price, chapter 10 of *Capital*, Vol. III (Marx [1894] 1981) already supplies the mechanism: value depreciates because social labor exceeded effective social want. Recovering the depreciated value, however, requires the new supply-demand equilibrium — precisely the demand-side degree of freedom left unmodelled in §5.2. Testing the two aggregate equalities at realized states is therefore not a convenient restriction:

socially necessary labor time is itself an equilibrium concept, requiring the minimal reproductive stability our Phase-0/Phase-1 states impose to be well defined at all.

**Wages.** The rising-wage scenario applies one economy-wide increment to the real-wage vector. This isolates distributional effects but excludes sectoral wage bargaining and labor-market heterogeneity. It should be read as an imposed wage scenario, not as an endogenous wage theory.

**Released capital.** Technical change devalues committed capital; whether the released value can be reinvested at the prevailing rate depends on accumulation outlets outside the adjustment rule.

**Money and fixed capital.** Money is treated as a labor-time numéraire, and credit and financial intermediation are excluded. All advanced capital is circulating capital. Fixed capital, historical-cost valuation and unequal turnover times could change both the admissible set and the transition and are left for later work.

*5.6 Structural Polarization: Capital Attraction and Innovation*

The objection that capital mobility contradicts observation turns on which profit rate governs migration. In values (unlike the equalized price rates of Figure 1A), sectoral rates at the pre-innovation equilibrium are 0.2635, 0.1122 and 0.1456 for bread, iron and meat, with organic compositions 3.26, 9.01 and 6.71; capital following those rates would desert capital-intensive branches. But competition responds to transformed profit rates. At equilibrium all three are 0.2033, and away from it migration follows deviations from that common rate. The apparent contradiction therefore arises only when the transformation is omitted (Salama and Valier 1973; Husson 2017).

Nor does capital invariably flow toward the branch with the highest organic composition. In Phase 0 it leaves capital-intensive Sector 2, whose profit rate is lowest; during diffusion Sector 2 becomes more profitable and attracts capital, though its committed capital still falls as devaluation exceeds the inflow. Attraction follows relative profitability, while committed capital depends on technical conditions; temporary pricing power over novel commodities can reinforce the former (Husson 2017; Laure van Bambeke 2018).[1]

Migration on the rate of Sector 2's new technique (Shaikh's 2016 regulating capital) leaves the final uniform rate unchanged in convergent runs and reproduces the pattern Shaikh describes: the new technique's rate is nearly equalized, while Sector 2's average rate, weighed down by older capital, falls below $\bar{r}$ during diffusion (Supplementary Section S4.7).

Husson (2017) also takes capital allocation as given. Yet once transfers of surplus value between branches are admitted, allocations compatible with uniform profit rates are constrained — uniquely in the two-branch case, as a one-parameter family here (the line of §5.1; Ankri, forthcoming). The identities delimit this set; competition selects within it.

The anchor selects different net products — (124.21, 63.95, 15.09) under bread anchoring, (126.47, 71.84, 4.94) under meat — implying the realization constraints (§5.5). Price competition can therefore affect real accumulation even when aggregate value is invariant, a route to structural polarization.

*5.7 Labor-Saving Innovation and Surplus Value*

The model assumes that living labor creates new value while material inputs transfer their cost. The simulation does not prove this axiom; it tests whether the resulting accounting remains coherent during diffusion. Sector 2's surplus value falls as direct labor declines, while its profit can remain above surplus value because the transformation redistributes surplus across sectors. A fully automated sector could therefore receive profit without creating surplus value, provided other sectors continue to generate it. Supplementary Section S4.3 gives the numerical example and accounting argument.

[1]This differs from Laure van Bambeke (2018; 2021), who preserves both equalities within an overdetermined system solved by least squares; the resulting solution satisfies them only approximately.

## 6. Conclusion

This paper constructs a dynamic three-sector model in which total price equals total value and total profit equals total surplus value at every recorded, fully realized state. Jointly, these equalities restrict compatible price-quantity configurations; an anchor selects a trajectory without prescribing firms' pricing intentions. Realization equilibrium at each step is distinct from the final equalization of profit rates. At uniform-profit equilibrium, the average rate coincides with the Perron–Frobenius rate of the corresponding Sraffian system. Along the transition, value accounting tracks the creation of surplus value and constrains its redistribution through prices and profits.

The analysis yields three narrower but robust findings. First, before innovation, capital inflow expands the receiving sector's output and, under both convergent anchors, lowers its realized unit price and excess profitability; at the initial allocation, the inflow narrows the gap between surplus value produced and the profit a uniform rate would require. Near equalization, general proofs establish that convergent paths exist, that stability is equivalent to a fall in the receiver's relative price, and that a stable approach lowers the receiver's profit rate even where its absolute price rises. The proofs are local; away from equalization, exceptions are rare and confined to extreme states (Supplementary Material S1.9). The direction of movement matches classical gravitation analyses, but the explanation differs: the fall in the receiver's rate results from the redistribution of surplus value required by the two aggregate equalities and the anchor, without any excess-supply price rule. Second, diffusion produces temporary extra surplus value for the new technique and releases capital through revaluation even while the innovating sector attracts capital on profitability grounds. Third, the final profit-rate response depends on distribution: it rises at constant real wages but falls when wage growth keeps aggregate exploitation approximately constant while organic composition increases.

These results neither prove value theory in general nor overturn Okishio's theorem. They support Marx's theory within the model's stated domain: a strict value-conservation specification is mathematically and computationally coherent, Okishio's result is recovered under its constant-real-wage assumption, and a different wage path can generate a falling rate after viable technical change. The main contribution is the transparent transition connecting the equilibria. The supplied code reproduces all 1,800 states, Figures 1–4 and the residual tests, letting readers vary the anchor, wage path, technical coefficients and adjustment speed.

**Data and Code Availability**

The complete Python reference implementation, figure-generation scripts, documentation, verification scripts and reference outputs are archived on Zenodo under the GNU GPL v3 license: https://doi.org/10.5281/zenodo.22933924. Running python value_conservation_simulation.py --run-all --batch reproduces the four scenarios; python generate_figures.py reproduces Figures 1–3 and python generate_figure4.py Figure 4; python supplementary_verification.py and python anchor_analysis.py reproduce the numerical checks in Supplementary Material S1, python relative_price_theorem.py those of the general results (Section S1.9), and python generate_figure_S1.py reproduces Figure S1 and python regulating_capital_rule.py and python anticipation_rule.py the variant migration rules of Section S4.7. The results were also reproduced by an independent LabVIEW implementation (§3.3), providing a cross-check against implementation error.

**Generative Artificial Intelligence Statement**

OpenAI's ChatGPT Plus and Anthropic's Claude Opus 5.5 were used for English-language editing, consistency checks, code review and preparation of submission materials. Claude Opus 5.5 also derived the proofs of the general results in Supplementary Material S1.9 (the fall of the receiver's profit rate, the existence of convergent paths and the relative-price theorem) from the authors' numerical findings and conjectures, and wrote the script that checks them. At the authors' request, it also implemented the variant migration rules of Supplementary Material S4.7. The authors designed the model, developed the theoretical argument, verified the calculations, the proofs and the code outputs, reviewed all revisions, and take full responsibility for the final content.

**Funding**

No funding was received for this work.

**Declaration of Conflicting Interests**

The authors declare no potential conflicts of interest with respect to the research, authorship and/or publication of this article.

# Supplementary Material

for the article *"On Dynamic Price Formation in the Course of Capital Reallocation Driven by Differential Rates of Profit: Strict Conservation of Value Supports Karl Marx's Theory"*
*by Norbert Ankri and Païkan Marcaggi*

*Reproduction code and reference outputs: Zenodo,*

*https://doi.org/10.5281/zenodo.22933924.*

## S1 Endogenous Price Response and Stability of Profit-Rate Equalization

This supplement derives and verifies the Phase-0 results summarized in Sections 3.1 and 5.2 of the paper. Each state assumes full realization of output; the sequence abstracts from the market fluctuations between states. The anchor selects a trajectory among configurations compatible with the two aggregate invariants, without specifying firms' pricing intentions. The derivations in Sections S1.2–S1.3 and the identities of Section S1.8 are algebraic within the model. Section S1.9 proves the stability, convergence and price results for every viable three-sector technique in which every commodity is basic; the simulations illustrate them. Unless stated otherwise, the numerical values in Sections S1.3–S1.10 are computed for the paper's technique ($A$, $\ell$ and $v$ of Section 2.1 of the paper) and would differ for another technique; the results proved in Sections S1.8–S1.9 hold for any technique that satisfies their assumptions. All tables and figures are computed from the authors' simulations.

### *S1.1 Fixed-technique notation*

The symbols used in Section S1 are listed below. Magnitudes measured in labor time and magnitudes measured at current prices are listed separately, since the argument turns on the relation between the two.

## Notation used in Section S1

| Symbol | Meaning |
| --- | --- |
| ***Indices*** | |
| i, j, m, n, k | Sectors: 1 bread, 2 iron, 3 meat. In Section S1.9, i and j are the two non-anchored sectors, labeled once and for all; in Section S1.9.5, k is a summation index (the unit capital $k_j$ is a different symbol) |
| a | Anchored sector, whose rate of profit equals the average rate $\bar{r}$ |
| h, l; d | Receiving and donating sector of a transfer; d marks the donor's rate $r_d$ (Section S1.9.1) |
| * | On q, κ and φ: the standard (reference) system. On z, K, r and p: equalization (z*, K* equalized allocations; r* uniform rate; p* production prices). r* is also the rate of the standard system at any prices (Section S1.8) |
| ***Technique and quantities*** | |
| A, ℓ, v | Input matrix, direct labor per unit of output, real-wage basket per hour of labor (Section 2.1 of the paper) |
| $\tilde{A} = A + v\ell$; $\tilde{a}_{nm}$ | Augmented sociotechnical matrix: means of production and wage goods per unit of output; its entry $\tilde{a}_{nm}$ is the quantity of commodity n used per unit of commodity m |
| $\lambda$; $p^*$, $q^*$ | Perron–Frobenius root of $\tilde{A}$, $0<\lambda<1$; left and right eigenvectors, $p^*\tilde{A}=\lambda p^*$ and $\tilde{A}q^*=\lambda q^*$. $q^*$ gives the reference production up to scale; $p^*$ gives production prices up to scale |
| $r^* = 1/\lambda - 1$ | Uniform rate of profit at production prices; at any prices, the rate of the standard system taken as a whole |
| $Q_j$; Q | Output of sector j; output vector, whose net product is $(I - \tilde{A})Q$ |
| I, $u_m$, **1** | Identity matrix, m-th unit vector, vector of ones |
| ***Magnitudes in labor time (values)*** | |
| $\Lambda_j$ | Unit value, $\Lambda = \ell(I - A)^{-1}$; $\Lambda v$ is the value of labor power per hour |
| $b_j$, $f_j$, $m_j$ | Value of the bread, iron and meat advanced per unit of output of sector j, wage goods included |
| $k_j$ | Unit capital advanced, $k_j = b_j + f_j + m_j$ (S1), that is, $k = \Lambda\tilde{A}$ |
| $s_j$; $\rho_j$ | Unit surplus value, $s_j = \Lambda_j - k_j$; value rate of profit, $\rho_j = s_j/k_j$ |
| $K_j$; $K_T$ | Capital advanced, $K_j = Q_j k_j$; total capital, $K_T = \Sigma_j K_j$ |
| $K_{init}$; K* | Initial capital allocation of Phase 0 in the paper; equalized allocation reached by the migration rule |
| $W_j$; $W_T$ | Value of output, $W_j = Q_j\Lambda_j$; its total |

| Symbol | Meaning |
|---|---|
| $B_j$, $F_j$, $M_j$ | Value of the bread, iron and meat advanced to sector j, $B_j = Q_j b_j$, $F_j = Q_j f_j$, $M_j = Q_j m_j$ |
| S | Total surplus value, $S = \Sigma_j \rho_j K_j$ |
| $\bar{r}$ | Average rate of profit, $\bar{r} = S/K_T = \Sigma_j \kappa_j \rho_j$ |
| $\kappa_j$; $\kappa^*_j$ | Actual capital shares, $\kappa_j = K_j/K_T$; shares of the standard system, $\kappa^*_j = k_j q^*_j/\Sigma_n k_n q^*_n$ |
| Γ | Gap between the profit a uniform rate would require and the surplus value produced, $\Gamma = r^* K_T - S$ |
| e; $OC_j$, OC_total, OC* | Rate of exploitation, uniform in Phase 0; organic composition of sector j and of the social capital, OC_total = C_total/V_total (Section 2.3 of the paper); that of the standard system, $OC^* = e/r^* - 1$ |
| ***Magnitudes at current prices*** | |
| $x_j$ | Transformation coefficient, $x_j = p_j/\Lambda_j$ |
| $p_j$ | Unit selling price, $p_j = \Lambda_j x_j$ (S3) |
| $c_j^p$; $c^p$ | Unit price-cost (S3); in vector form $c^p = p\tilde{A}$ |
| $C_j^p$ | Capital advanced at current prices, $C_j^p = Q_j c_j^p$; $\Sigma_j C_j^p = K_T$ by equations (1) and (2) of the paper |
| $\Pi_j$ | Profit, $\Pi_j = W_j x_j - C_j^p$; $\Sigma_j \Pi_j = S$ by equation (2) of the paper |
| $r_j$ | Rate of profit, $1 + r_j = p_j/c_j^p$ (S4) |
| $\varphi_j$; $\varphi^*_j$ | Capital shares at current prices: actual, $\varphi_j = C_j^p/\Sigma_n C_n^p$; of the standard system, $\varphi^*_j = q^*_j c_j^p/\Sigma_n q^*_n c_n^p$ |
| $\sigma_j$ | Representation of sector j in the standard system relative to the actual economy, $\sigma_j = \varphi^*_j/\varphi_j$ |
| $\Psi$; $\Psi_{mn}$ | Input cost-share matrix, $\Psi_{mn} = p_n \tilde{a}_{nm}/c_m^p$; each row sums to one (Section S1.9.5) |
| $\tau_{mn}$; $\tau^{(n)}$ | Mean number of input links followed from $m$ until the first encounter with $n$, weighted by cost shares, with $\tau_{nn} = 0$; $\tau^{(n)} = (\tau_{mn})_m$ is the vector for that target (Section S1.9.5) |
| $\omega_m$ | Share of total input expenditure devoted to commodity $m$, $\omega_m = p_m (\tilde{A} Q)_m / K_T = \sum_n \varphi_n \Psi_{nm}$ (Section S1.9.5) |
| $\xi_m$; $\xi$; $r'$ | At $z^*$, $\xi_m = d \ln p_m / dz$, $\xi = (\xi_m)_m$ and $r' = (r'_m)_m$; log-price and profit-rate derivative vectors (Section S1.9.5) |
| $\xi^{(0)}$ | Particular solution of $(I - \Psi)\xi = r'/(1 + r^*)$, constructed from first-passage times (Section S1.9.5) |

| Symbol | Meaning |
|---|---|
| $\zeta$ | Common component in $\xi=\xi^{(0)}+\zeta 1$, determined by the aggregate price-level constraint (Section S1.9.5) |
| $\chi$ | Distance between actual and reference capital shares, $\chi^2=\sum_j (\varphi_j^*-\varphi_j)^2/\varphi_j$; enters the bound on profit-rate dispersion (Section S1.10) |
| ***Adjustment path*** | |
| z; z* | Capital of the receiving sector h (Sections S1.2–S1.3) or of sector i (Section S1.9), the anchored capital being fixed; an equalized allocation |
| $z_s$ | Allocation at which the fixed-anchor segment intersects the singular line of the anchor (Section S1.9.3) |
| t | Index of successive states under the migration rule (Section S1.3) |
| $\delta_t$ | Deviation from the equalized allocation at step $t$, $\delta_t=z_t-z^*$ (Section S1.3) |
| g(z) | Profit-rate gap between the two non-anchored sectors (S6): $r_h - r_l$ in Sections S1.2–S1.3, $r_i - r_j$ in Section S1.9 |
| α; μ | Migration coefficient; local multiplier μ = 1 + (α/r*)g′(z*) (S7) |
| $\alpha_{crit}$ | Critical migration coefficient, $\alpha_{crit}=-2r^*/g'(z^*)$, when $g'(z^*)<0$ (Section S1.4) |
| T(z) | Migration map, $T(z) = z + \alpha g(z)/r_d(z)$ (Section S1.9.1) |
| J | Interval of starting allocations attracted by z* (Proposition 3) |
| $H(z)$, $y(z)$ | Coefficient matrix and right-hand-side column vector in $H(z)x(z)=y(z)$ (Section S1.2) |
| $\eta$; $Y$ | Auxiliary row vectors in the regularity proof: $\eta$ solves the homogeneous price system at $z^*$; $Y=\eta[I-(1+r^*)\widetilde{A}]$ (Section S1.9.3) |
| E; D | $E = \kappa^*_i\kappa_j - \kappa^*_j\kappa_i$, zero on the singular line of anchor a; $D = \varphi^*_i\varphi_j - \varphi^*_j\varphi_i$, of the same sign (Section S1.9.3) |
| $v$ | Scalar $v=-g'/(\varphi_i+\varphi_j)$ at $z^*$, used to express profit-rate derivatives (Sections S1.9.5–S1.9.6) |
| ′ | Derivative with respect to z, at z* unless stated otherwise. In the paper, a prime marks instead the post-innovation technique (A′, ℓ′, Λ′) |

During Phase 0, A, ℓ, v and Λ are fixed. For sector i, the unit capital advanced in value is:

$$k_i=b_i+f_i+m_i \tag{S1}$$

Since $K_i = Q_i k_i$, sectoral output satisfies:

$$Q_i = \frac{K_i}{k_i}, \frac{dQ_i}{dK_i} = \frac{1}{k_i} > 0 \tag{S2}$$

Let $x = (x_1, x_2, x_3)$ be the transformation coefficients. The unit selling price and the unit price-cost of sector i are:

$$p_i = \Lambda_i x_i, c_i^p = b_i x_1 + f_i x_2 + m_i x_3 \tag{S3}$$

The corresponding aggregate price-cost is

$$C_i^p = Q_i c_i^p = B_i x_1 + F_i x_2 + M_i x_3.$$

Hence:

$$1 + r_i = \frac{W_i x_i}{C_i^p} = \frac{\Lambda_i x_i}{b_i x_1 + f_i x_2 + m_i x_3} = \frac{p_i}{c_i^p} \tag{S4}$$

Equation (S4) contains no $K_i$ or $Q_i$. Therefore an inflow of capital, by itself and at fixed relative prices, expands output, aggregate price-cost and the mass of profit proportionally but cannot change the sectoral profit rate. Any equalizing feedback must operate through the endogenous response of x to the new capital allocation.

*S1.2 Implicit price response to capital reallocation*

At each Phase-0 state, equations (1)–(3) of the main paper determine the transformation-coefficient vector from the current aggregates and the anchor condition. For compactness, write this system as

$$H(z)\, x(z) = y(z),$$

where z is the capital committed to one non-anchored sector, $x(z) = (x_1(z), x_2(z), x_3(z))^T$ is the vector of transformation coefficients, H(z) is the 3 × 3 coefficient matrix obtained by collecting the coefficients of $x_1$, $x_2$ and $x_3$ in these three equations, and y(z) is the corresponding right-hand-side vector.

For an anchor in sector a:

$$y(z)=(S(z),W_1(z)+W_2(z)+W_3(z),0)^T$$

The first component corresponds to the equality between total profit and total surplus value, the second to the equality between total price and total value, and the third to the anchored-sector price equation written with all terms involving x on the left-hand side. Both H(z) and y(z) depend on z through the current sectoral capital allocation and the resulting aggregates.

The matrix H(z) is only an auxiliary matrix notation for the contemporaneous price system; it is not a technological matrix. In particular, it is distinct from the augmented sociotechnical matrix $\tilde{A} = A + v\ell$ of the main paper. Whereas $\tilde{A}$ describes unit physical reproduction conditions and remains fixed throughout Phase 0, H(z) collects the aggregate coefficients entering the price equations and therefore changes when capital is reallocated. Where H(z) is invertible, differentiation with respect to z (denoted by a prime) gives, by the product rule,

$$H'(z)x(z)+H(z)x'(z)=y'(z).$$

Moving the first term to the right and multiplying on the left by $H(z)^{-1}$ gives

$$x'(z)=H(z)^{-1}[y'(z)-H'(z)x(z)] \quad \text{(S5)}$$

With the anchored sector fixed, the two non-anchored capitals sum to a constant. If h is the receiving sector and l the donating sector, set $z = K_h$ and $K_l$ = constant − z. Because $Q_h'(z) = 1/k_h > 0$, an inverse quantity-price response is equivalent to $p_h'(z) < 0$.

*S1.3 Local stability of profit-rate equalization*

Define the profit-rate gap between the two non-anchored sectors by:

$$g(z) = r_h(z) - r_l(z) \tag{S6}$$

At the uniform-profit allocation z*, g(z*) = 0 and $r_h$(z*) = $r_l$(z*) = r*. When h is the more profitable sector, the migration rule of the paper is

$$z_{t+1} = z_t + \alpha \frac{g(z_t)}{r_l(z_t)}.$$

Writing $z_t = z^* + \delta_t$, define $T(z) = z + \alpha g(z)/r_l(z)$ on this side of the equilibrium. Since $g(z^*) = 0$ and $r_l(z^*) = r^*$, differentiation gives

$$T'(z^*) = 1 + \alpha \left[ \frac{g'}{r_l} - \frac{g\, r_l'}{r_l^2} \right]_{z=z^*} = 1 + \frac{\alpha}{r^*} g'(z^*).$$

Taylor expansion of $z_{t+1} = T(z^* + \delta_t)$, followed by subtraction of $z^* = T(z^*)$, therefore gives

$$\delta_{t+1} = z_{t+1} - z^* = T(z^* + \delta_t) - T(z^*).$$

Taylor's formula makes this difference $T'(z^*)\delta_t + o(\delta_t)$. Substituting the derivative above gives the first-order relation

$$\delta_{t+1} = \left[ 1 + \frac{\alpha}{r^*} g'(z^*) \right] \delta_t \tag{S7}$$

The remainder is $o(\delta_t)$; (S7) records the first-order dynamics.

Strict attraction of the linearization, which guarantees local asymptotic stability, requires the multiplier to have modulus below unity:

$$\frac{-2r^*}{\alpha} < g'(z^*) < 0 \tag{S8}$$

The economic part of (S8), $g'(z^*) < 0$, states that an inflow of capital into the more profitable sector must reduce its profitability advantage. The lower bound prevents adjustment from being so strong that successive overshooting becomes explosive. More specifically, $-r^*/\alpha < g'(z^*) < 0$ gives monotone convergence, while $-2r^*/\alpha < g'(z^*) < -r^*/\alpha$ gives damped oscillatory convergence. Indeed, the first range gives $0<\mu<1$: the deviation keeps its sign and shrinks. The second gives $-1<\mu<0$: its sign alternates while its magnitude shrinks. The boundary cases $\mu=\pm 1$ require higher-order analysis.

Condition (S8) has a direct economic reading. Let h index the most profitable branch, the one receiving capital. Writing its profit as $\Pi_h$ and its committed capital as $C_h^p$, both measured in price, we have

$$r_h = \frac{\Pi_h}{C_h^p}, \text{hence} \frac{d r_h}{d C_h^p} = \frac{1}{C_h^p} \frac{d \Pi_h}{d C_h^p} - \frac{\Pi_h}{\left(C_h^p\right)^2}.$$

The local convergence requirement $dr_h/dC_h^p < 0$ is therefore exactly equivalent to

$$\frac{d \Pi_h}{d C_h^p} < \frac{\Pi_h}{C_h^p}:$$

additional capital must raise the branch's profit less than proportionally. Measuring both magnitudes in price avoids any approximation; the same sign follows if the derivative is taken with respect to capital in value, since $dC_h^p/dz > 0$ at both equalized allocations. This is a branch-level convergence criterion. It does not express diminishing returns: the technique has constant returns ($Q_h = K_h/k_h$), and profit grows less than proportionally only through the response of relative prices to the reallocation. Its failure describes a branch whose appropriated profit grows faster than the capital committed to it, so that competition concentrates capital rather than equalizing profit rates. The criterion was first stated by Ankri and Marcaggi (2022, Section D), with capital measured in value and profit-rate differentials prescribed rather than selected by an anchor. Both convergent anchors satisfy it with a wide margin: $d\Pi_h/dC_h^p = 0.033$ against $\Pi_h/C_h^p = 0.203$ under bread anchoring, and −3.87 against 0.203 under meat. Anchoring Sector 2 violates it at the same uniform-profit allocation: there $d\Pi_h/dC_h^p = 1.207$ exceeds $\Pi_h/C_h^p = 0.203$ and $dr_h/dz = +1.84 \times 10^{-3}$, so the equilibrium is repelling. A displacement of 0.5 units then drains the donating sector from $K_3 = 188.61$ to 2.87 within forty transfers, at constant KT, until no economically admissible price system remains. Well before then, meat output stops covering the meat that the three branches use as input and wage good (its net product turns negative), so the path also violates physical reproduction. The failure is a property of the closure at a given allocation rather than of the sector (Section S1.9): at the same allocation and the same technique, Sector 1 satisfies the criterion when Sector 3 is anchored ($d\Pi_h/dC_h^p = -10.55$ against $\Pi_h/C_h^p = 0.203$) and violates it when Sector 2 is anchored (1.207 against 0.203). Note finally that $g' = r_h' - r_l'$ involves both branches: at the meat-anchored equilibrium $r_l'(z^*) = +7.39 \times 10^{-3}$ exceeds $|r_h'(z^*)| = 3.27 \times 10^{-3}$. At an equalized allocation the two conditions are nevertheless equivalent: under the anchor closure, $g'(z^*) < 0$ if and only if the receiver's rate falls as it receives capital, from either side (Section S1.9: Proposition 1 and the proof of part (i) of the theorem).

### *S1.4 Numerical values at the two Phase-0 equilibria*

Implicit differentiation of the calibrated price system gives the following derivatives per unit of capital. The multiplier μ is evaluated at $\alpha = 10$ and $\alpha crit = -2r^*/g'$ is the local critical migration coefficient.

| Anchor | Active transfer | Limiting K* | g′(z*) | $p_h'(z^*)$ | μ (α=10) | αcrit |
|---|---|---|---|---|---|---|
| Bread (1) | 2 → 3 | (560, 251.3904146, 188.6095854) | $-1.6744\times10^{-3}$ | $-1.1293\times10^{-4}$ | 0.9176 | 242.8 |
| Meat (3) | 2 → 1 | (561.9018018, 258.0981982, 180) | $-1.0665\times10^{-2}$ | $-5.2921\times10^{-4}$ | 0.4753 | 38.1 |

Both equilibria satisfy (S8). The much smaller multiplier under meat anchoring explains why the main simulation reaches its migration tolerance much sooner with that closure. The Phase-0 tests reported in this supplement use an extended horizon, run until the $10^{-10}$ tolerance is met (iteration 238 under bread anchoring, 30 under meat), so that the fixed point is reached to $10^{-10}$; the main text reports the standard Phase 0, which ends at iteration 200 when innovation begins. The two K* values therefore differ in the sixth decimal.

*S1.5 Iteration-by-iteration sequence test*

The reference implementation uses the migration stopping criterion $\max_j|r_j - \bar{r}|/\bar{r} \leq 10^{-10}$. Before that criterion is reached, every effective Phase-0 transfer was tested. For the receiving sector h we checked $\Delta K_h > 0$, $\Delta Q_h > 0$, $\Delta p_h < 0$, $\Delta(p_h/c_h^p) < 0$ and $\Delta r_h < 0$; for the donating sector l the price and profit-rate movements have the opposite sign. The absolute profit-rate gap also contracts after each transfer.

| **Anchor** | **Effective transfers** | **Receiving sector** | **$\Delta K_h$** | **$\Delta Q_h$** | **$\Delta p_h$** | **$\Delta C_h^p$** | **$\Delta \Pi_h$** | **$r_h$ start → stop** |
|---|---|---|---|---|---|---|---|---|
| Bread | 236 / 236 | 3 (meat) | +4.783% | +4.783% | −0.388% | +4.864% | +2.039% | 0.208879 → 0.203251 |
| Meat | 28 / 28 | 1 (bread) | +0.340% | +0.340% | −0.619% | +0.175% | −2.468% | 0.208760 → 0.203251 |

The bread-anchor case also illustrates that $r_h$ can fall while the sector's mass of profit $\Pi_h$ still rises: what matters is that $\Pi_h$ grows more slowly than $C_h^p$. Thus $r_i$ ↓ does not imply $\Pi_i$ ↓.

*S1.6 Robustness along the calibrated allocation branch and role of α*

A dense numerical scan over the economically admissible positive-capital branch was performed with the paper's technology and wage basket. With bread anchoring, $K_1$ is fixed and $K_2 + K_3 = 440$; with meat anchoring, $K_3$ is fixed and $K_1 + K_2 = 820$. Wherever the price system remains economically admissible, the scan gives both $p_h'(z) < 0$ and $g'(z) < 0$ for the relevant non-anchored pair. Hence, for this parameterization, the inverse quantity-price relation is not restricted to the realized trajectory between the initial state and equilibrium.

The local stability threshold is not a global convergence guarantee. Simulations nevertheless behave as predicted near the threshold: with bread anchoring $\alpha = 200$ reaches the $10^{-6}$ dispersion level, $\alpha = 240$ remains very slowly oscillatory over the Phase-0 horizon, and $\alpha = 245$ leaves a much larger residual dispersion; with meat anchoring $\alpha = 35$ converges, whereas $\alpha = 38$, although just below the local $\alpha_{crit}$, does not enter the local basin from the paper's initial allocation within the Phase-0 horizon. This distinction between local stability and global attraction is important.

The relevant range of α is the low one. With total capital fixed, reallocation can only move existing capital between sectors, and in actual economies only a small fraction of a sector's capital changes sector from one period to the next. The paper's $\alpha = 10$ respects this: at $K_{init}$, the first transfer moves 0.50 units of capital under the bread anchor and 0.94 under the meat anchor, that is, 0.19–0.36 percent of the donor's capital and 0.17–0.28 percent of the receiver's. Such steps lie within the small-α regime assumed by the local results of Sections S1.3 and S1.9. They do not, however, keep the economy close to equalization: a shock that displaces the equalized allocation, such as the innovation of Phase 1, can leave the state far from the new target even when capital moves slowly.

*S1.7 Interpretation of the sequence of realized states*

The verified adjustment under the two convergent closures can be summarized as:

$$r_h > r_l \rightarrow \text{capital reallocation} \rightarrow \text{new Q and S} \rightarrow \text{compatible } x \rightarrow p_h \downarrow \text{and } g \downarrow \qquad \text{(S9)}$$

Capital reallocation expands $Q_h$ and contracts $Q_l$ by (S2), while changing aggregate surplus value and the coefficients of the conservation equations. Equations (1)–(3) then select the price vector of the next state. Each recorded state assumes that supply and demand have met and output has sold, but sectoral profit rates can remain unequal. The sequence therefore concerns realization equilibria before uniform-profit equilibrium. It does not model the negotiations, expectations or price fluctuations that clear markets, and (S9) does not assert a direct excess-supply price rule. The fall in $p_h$ is a verified response of the selected compatible states.

The result is an inverse quantity-price response along the calibrated sequence of realized states, conditional on value conservation and the closure. The invariants restrict the possible configurations; the anchor selects one trajectory; it is not a behavioral target for firms. Section S1.9 proves, for every viable three-sector technique in which every commodity is basic, that near equalization a stable approach lowers the receiver's profit rate. It also gives a necessary and sufficient stability condition, shows that convergent paths exist, and shows that stability is equivalent to a fall in the receiver's price relative to the donor's, its absolute price falling from at least one side. The fall of $p_h$ itself, verified along the stable paths of the present technique, does not hold for every technique: it can fail on one side of z* and away from equalization (Section S1.9). These calculations do not identify a demand function.

*S1.8 Mechanism of the price response: surplus-value gap and standard system*

The price response rests on two averages of the same sectoral rates (S4). The first weights each rate by the sector's share in the capital actually advanced, valued at current prices,

$$\varphi_j = \frac{Q_j c_j^p}{\sum_i Q_i c_i^p}.$$

Because equations (1) and (2) of the paper imply $\Sigma_j C_j^p = K_T$ and $\Sigma_j \Pi_j = S$, this average equals $\bar{r}$ at every state:

$$\sum_j \varphi_j r_j = \frac{\sum_j \Pi_j}{\sum_j C_j^p} = \frac{S}{K_T} = \bar{r}.$$

The second weights the same rates by the capital shares of a reference production. Let $q^*$ be the right Perron–Frobenius eigenvector of $\tilde{A}$, positive because every commodity is basic (Section S1.9.1), so that $\tilde{A} q^* = \lambda q^*$ with $\lambda = 1/(1 + r^*)$, and read its components as quantities produced. Producing $q^*$ requires inputs, wage goods included, equal to $\lambda q^*$: gross output and advances are the same bundle of commodities in different amounts. The physical surplus is $q^* - \tilde{A} q^* = r^* \tilde{A} q^*$. Valued at any positive prices $p$, the capital of this reference production is

$$\sum_j q_j^* c_j^p = p \tilde{A} q^* = \lambda p q^*$$

and its profit is (1 − λ)pq*, so that its aggregate rate of profit is (1 − λ)/λ = r* whatever the prices: the prices cancel because they value two physically proportional bundles. Let

$$\varphi_j^* = \frac{q_j^* c_j^p}{\sum_i q_i^* c_i^p}$$

be the reference capital shares at current prices. Since

$$\varphi_j^* r_j = \frac{q_j^* \left( p_j - c_j^p \right)}{\sum_i q_i^* c_i^p},$$

the φ*-weighted average of the sectoral rates is the aggregate rate of the reference production, r*, at every state, including states in which the rates differ:

$$\sum_j \varphi_j^* r_j = r^*.$$

The scale of $q^*$ is arbitrary: $\sum_n k_n q_n^*$ is the reference capital in value and need not equal $K_T$, although that normalization may be chosen. Rescaling $q^*$ leaves its capital shares unchanged; the effective outputs remain $Q_m = K_m / k_m$.

For example, with Ã = ((0.4, 0.2), (0.2, 0.7)) and q* = (100, 200), λ = 0.8 and r* = 0.25; at prices (1, 1) the two rates are 2/3 and 1/9, the weights 1/4 and 3/4, and the weighted average 1/4. The weights $\varphi^*_j$ vary with prices and differ from the actual shares $\varphi_j$, and the reference proportions are not, in general, those of the equalized allocation actually reached.

This reference production is Sraffa's (1960, chs. IV–V) standard system, here built on the augmented matrix $\tilde{A}$, so that wage goods count among the means of production and r* is the uniform rate at the given real wage, not Sraffa's maximum rate of profit. Sraffa used the standard commodity as an invariable measure of value, in terms of which the relation between the wage and the rate of profit becomes linear and independent of prices (see Bellino 2004). It serves a different purpose here: its physical proportionality yields an identity that constrains the movement of the sectoral rates. No commodity is taken as numéraire, prices remain those selected by equations (1)–(3) of the paper, and the identity adds no equation to the closure. In the standard system each commodity is produced in the same proportion 1 + r* = 1.2033 to its total use as means of production and wage good, whereas at $K_{init}$ the ratios $R_j$ of output to total use are (1.2133, 1.3411, 1.0249). Off equilibrium, the sectors of the standard system earn unequal rates, but their profits in excess of remuneration at r* cancel out, as transfers do in Marx's aggregate accounting. The rate r* thus plays two roles: the uniform rate at production prices and, at any prices, the rate of the standard system taken as a whole; identity (S10) uses only the second. On the standard commodity and Marx's value categories, see Eatwell (1975) and, critically, Sinha (2000). Subtracting the two averages gives:

$$\sum_j \left(\varphi_j^* - \varphi_j\right) r_j = r^* - \bar{r} \tag{S10}$$

Identity (S10) uses only equations (1) and (2), not the anchor: it holds at every state and for any rule that selects prices within the admissible set (numerically, its largest residual is below $2 \times 10^{-15}$). It has three consequences.

First, whatever the rule that selects prices, uniform rates are possible only if $\bar{r} = r^*$, that is, if the surplus value produced equals the profit that a uniform rate would require:

$$S = r^* K_T .$$

Since $S = \Sigma_j \rho_j K_j$ with $\rho_j = s_j/k_j$, this condition is the straight line

$$\sum_j \kappa_j \rho_j = r^*$$

in the simplex of capital shares $\kappa_j = K_j/K_T$, and the line contains the standard proportions

$$\kappa_j^* = \frac{q_j^* k_j}{\sum_i q_i^* k_i},$$

here κ* = (0.5552, 0.2345, 0.2103) (Section S1.9.3). Without fixed capital, Ankri and Marcaggi (2022, Section C-1) already described this set as a line along which total surplus value is constant. Laure van Bambeke (2021, 181) describes a straight line of solutions in a model where the depreciation of fixed capital is not transformed; the set is then slightly curved (Ankri and Marcaggi 2022, Section B-1). The equalized allocations reached in the paper, (0.5600, 0.2514, 0.1886) under the bread anchor and (0.5619, 0.2581, 0.1800) under the meat anchor, and the allocation (0.5539, 0.2297, 0.2164) of Section S1.9 all give $\Sigma_j \kappa_j \rho_j$ = 0.20325 = r*, to the rounding of the shares. Conservation thus fixes the possible terminal states, allocations on this line with Sraffa's prices up to scale, while the selection rule decides which of them is reached, and whether one is reached at all.

In Phase 0 all sectors share the rate of exploitation e = 1.1236, so that $\rho_j = e/(1 + OC_j)$, with $OC_j$ the organic composition of Section 2.3 of the paper (3.264, 9.012 and 6.714 for bread, iron and meat), and $\bar{r} = e/(1 + OC_total)$ at every state, as for Marx's general rate (Marx [1894] 1981, ch. 9). The line of equalized allocations is therefore the locus on which the social capital has the organic composition of the standard system,

$$\mathrm{OC}^{*}=\frac{e}{r^{*}}-1=4.528\,,$$

and $r^* - \bar{r}$ measures the departure of the social composition from that value (OC_total = 4.536 at $K_{init}$). Uniform rates are therefore possible only where the value rate of the social capital, S/(C_total + V_total), equals the uniform rate. This bears on Steedman's (1977) objection that the two generally differ, an objection that concerns a given allocation: under the conservation hypothesis, allocations off this line admit no uniform rate. Two qualifications apply. At the end of the process it is r*, fixed by the technique and the real wage, that determines the equilibrium social composition, not the reverse. And the reading requires a uniform rate of exploitation, so it does not extend to the diffusion phase, where $e_j$ differs across sectors.

Second, let

$$\Gamma=r^{*}K_{T}-S=\sum_{j}\left(r^{*}-\rho_{j}\right)K_{j}$$

be the gap between the profit consistent with a uniform rate and the surplus value produced, and let $\sigma_j = \varphi^*_j/\varphi_j$. For each sector, $C_j^p(r_j - \bar{r})$ is the difference between its profit, $C_j^p r_j$, and the profit its capital would earn at the current average rate, $C_j^p\bar{r}$. The actual average, which weights the rates by the capital valued at current prices, equals $\bar{r}$ at every state (see above):

$$\bar{r} = \frac{C_a^p r_a + C_h^p r_h + C_l^p r_l}{C_a^p + C_h^p + C_l^p}.$$

Multiplying by the denominator and collecting terms gives

$$C_a^p(r_a - \bar{r}) + C_h^p(r_h - \bar{r}) + C_l^p(r_l - \bar{r}) = 0.$$

The anchor makes the first term vanish, since $r_a = \bar{r}$. With sector a anchored at $\bar{r}$, the two other sectors' excess profits, measured relative to remuneration of their capital at the current average, therefore cancel:

$$C_h^p(r_h - \bar{r}) + C_l^p(r_l - \bar{r}) = 0.$$

The reference weights also sum to one, so that subtracting $\bar{r}$ from every rate in the reference average gives $\Sigma_j\varphi^*_j(r_j - \bar{r}) = r^* - \bar{r}$. The anchored term again vanishes, and $r^* - \bar{r} = (r^*K_T - S)/K_T = \Gamma/K_T$, whence

$$\varphi_h^*(r_h - \bar{r}) + \varphi_l^*(r_l - \bar{r}) = \frac{\Gamma}{K_T}.$$

Since $\varphi_j = C_j^p/K_T$, the reference weights are $\varphi^*_j = \sigma_j C_j^p/K_T$. Substituting them into this equation, multiplying by $K_T$ and replacing $C_l^p(r_l - \bar{r})$ by $-C_h^p(r_h - \bar{r})$ gives

$$(\sigma_h - \sigma_l) C_h^p (r_h - \bar{r}) = \Gamma,$$

so that, wherever $\sigma_h \neq \sigma_l$,

$$(r_h - \bar{r}) C_h^p = -(r_l - \bar{r}) C_l^p = \frac{\Gamma}{\sigma_h - \sigma_l} \qquad \text{(S11)}$$

These equal and opposite deviations from remuneration at the average rate are proportional to Γ and inversely proportional to the difference in representation relative to the standard system. They are not the differences between each branch's profit and its own produced surplus value; those differences generally remain non-zero at uniform rates. Because the unit costs cancel in $\varphi^*_j/\varphi_j$,

$$\sigma_h - \sigma_l = \left( \frac{q_h^*}{Q_h} - \frac{q_l^*}{Q_l} \right) \frac{\sum_i Q_i c_i^p}{\sum_i q_i^* c_i^p}$$

and $C_h^p$ and $C_l^p$ are positive, $r_h - \bar{r}$ and $r_l - \bar{r}$ have opposite signs and the rate ranking satisfies

$$\operatorname{sign}(r_h - r_l) = \operatorname{sign}(\Gamma) \cdot \operatorname{sign}\left( \frac{q_h^*}{Q_h} - \frac{q_l^*}{Q_l} \right).$$

When Γ > 0, the sector lagging further behind its standard size earns the premium; when Γ < 0, the sector further ahead does. As $\sigma_h$ approaches $\sigma_l$, the excess profits must grow unless Γ vanishes first: the line through vertex a and κ* is singular for anchor a: where a segment of anchor a crosses it, the selected prices give the standard system a zero cost, so that its weights are undefined, and prices become inadmissible near it.

The bread-anchored trajectory gives a concrete accounting example. Between $K_{init}$ and uniform-profit equilibrium, meat output and its own surplus value grow by 4.783 percent, but its appropriated profit grows by 2.039 percent (Section S1.5). The unit cost rises by about 0.078 percent, while profit per unit falls sufficiently to lower the selling price by 0.388 percent. The price is the sum of unit cost and profit per unit; no unsold output is required in this comparison of realized states. Meat's excess profit relative to the current average falls from about 1.079 to zero. Its profit minus its own surplus value, however, falls only from 11.890 to 11.413: ordinary redistribution of surplus value persists after the profitability premium disappears.

Third, the receiver's premium

$$r_h - \bar{r} = \frac{\Gamma}{\left(\sigma_h - \sigma_l\right) C_h^p}$$

responds to a transfer through three terms: the gap Γ, the distance to standard proportions $\sigma_h - \sigma_l$ and the receiver's capital $C_h^p$. Table S5 reports their log-derivatives at $K_{init}$, where the gap dominates in all five cases. Away from $K_{init}$ the proportion term can dominate: from shares (0.55, 0.26, 0.19) under the bread anchor, a transfer from iron to meat narrows Γ by 2.3 percent but raises the premium by 1.7 percent, and the adjustment diverges until iron's capital is exhausted. Each unit of capital carries its own value rate of profit, so that

$$d\Gamma = \sum_j \left(r^* - \rho_j\right) d K_j :$$

a transfer from iron ($\rho_2 = 0.112$) to meat ($\rho_3 = 0.146$) lowers the social organic composition, raising S and narrowing Γ, whereas a transfer from bread ($\rho_1 = 0.264$) to meat, or new capital placed in meat, widens it. In terms of living labor, one unit of capital sets 0.498 hours in motion in bread, 0.275 in meat and 0.212 in iron; at $K_{init}$ the social capital sets 383.61 hours in motion, whereas uniform rates would require $r^*K_T/(1 - \Lambda v) = 384.15$, so that $\Gamma = (1 - \Lambda v)(384.15 - 383.61) = 0.29$.

**Table S5 Log-derivatives at $K_{init}$ per unit of capital moved (percent) and change in the receiver's transformation coefficient**

| Anchor | Movement | Gap Γ | Proportion | Capital | Premium | $\Delta x_h$ per unit |
|---|---|---|---|---|---|---|
| Bread | iron → meat | −11.61 | +3.74 | −0.56 | −8.44 | $-3.46 \times 10^{-4}$ |
| Meat | iron → bread | −52.58 | +5.85 | −0.10 | −46.83 | $-2.73 \times 10^{-3}$ |
| Iron | bread → meat | +40.97 | +4.66 | −0.55 | +45.08 | $+3.37 \times 10^{-3}$ |
| Bread | new capital → meat | +20.02 | +2.33 | −0.54 | +21.82 | $+8.98 \times 10^{-4}$ |
| Bread | bread → meat | +40.97 | +2.43 | −0.52 | +42.88 | $+1.76 \times 10^{-3}$ |

Note: the premium is $r_h - \bar{r} = \Gamma/[(\sigma_h - \sigma_l)C_h^p]$; Proportion and Capital are the log-derivatives of $1/(\sigma_h - \sigma_l)$ and $1/C_h^p$, and the three middle columns add up to the premium column up to rounding. The pair (h, l) is (meat, iron) under the bread anchor, (bread, iron) under the meat anchor and (meat, bread) under the iron anchor. Source: Authors' simulations.

*S1.9 General results: stability, convergence and the price response*

This section proves, for every viable three-sector technique in which every commodity is basic, general counterparts of the local results that Sections S1.3–S1.7 establish numerically for the paper's technique; for that technique, the simulations, Table S6 and Figure S1 illustrate them (Section S1.9.7). Proposition 1 shows that near a stable equalized allocation each transfer lowers the receiver's profit rate and raises the donor's, whatever happens to the receiver's selling price. Proposition 2 gives a necessary and sufficient condition for stability in terms of technical and value data alone; the same condition fixes the direction of every transfer along the path. Proposition 3 shows that convergent paths exist: for small enough α, every equalized allocation that satisfies the condition attracts an interval of starting allocations, and such allocations exist for every technique whose value rates of profit are not all equal. A theorem then relates stability to the receiver's price relative to the donor's. The proofs use the notation of Sections S1.1, S1.3 and S1.8 and are otherwise self-contained. Apart from the direction of transfers, the results are local; Section S1.9.8 reports what happens away from equalization.

The following table lists these results and those of Section S1.8, with the section in which each is established and the steps on which it relies.

**Results of Sections S1.8–S1.9 and the steps on which they rely**

| **Result** | **Section** | **Relies on** |
|---|---|---|
| Actual average, $\Sigma_j\varphi_j r_j = \bar{r}$ | S1.8 | Equalities (1) and (2) of the paper, through $\Sigma_j C_j^p = K_T$ and $\Sigma_j \Pi_j = S$ |
| Reference average, $\Sigma_j\varphi^*_j r_j = r^*$ | S1.8 | Physical proportionality of the standard system, $\tilde{A}q^* = \lambda q^*$; holds at any prices |
| Identity (S10) | S1.8 | The two averages; not the anchor |
| Line of equalized allocations, $\Sigma_j\kappa_j\rho_j = r^*$ | S1.8, S1.9.3 | (S10) with uniform rates (necessity); prices proportional to p* (sufficiency) |
| Excess profits (S11) | S1.8 | Actual average and anchor (the excess profits cancel); reference average (their size); three sectors |
| Proposition 1: the receiver's rate falls, the donor's rises | S1.9.2 | Actual average and anchor (ordering of the derivatives); reference average (their signs) |
| Proposition 2: regularity, condition (S12), formula (S13) | S1.9.3 | Regularity: λ simple, and q*, Q and $u_a$ independent off the singular line; (S13): $\bar{r}' = (\rho_i - \rho_j)/K_T$ and steps (a)–(b) of Proposition 1; sign: geometry of the line of equalized allocations |
| Direction of transfers along the segment | S1.9.3 | (S11) and the sign of $g'(z^*)$ (Proposition 2) |
| A stabilizing anchor exists | S1.9.3 | $\Sigma_m(\kappa_m - \kappa^*_m)(\rho_m - r^*) = 0$ and (S12) |
| Proposition 3: convergent paths exist | S1.9.4 | Propositions 1 and 2, the formula for E (step (c) of the proof of Proposition 2) and a contraction argument |
| Theorem (i): sign of the relative price | S1.9.5 | Signs of $r_i'$ and $r_j'$ (Proposition 1); unit-cost changes are averages of input-price changes |
| Theorem (ii), (S14): size of the response | S1.9.5 | Step (b) of the proof of Proposition 2 and the cost-share Markov chain Ψ (first-passage times) |
| Extension to n sectors | S1.9.6 | The two averages and the first-passage identity |
| Any change in capital | S1.9.7 | Proof of Proposition 1 and (S11) |

*S1.9.1 Assumptions, notation and the migration map*

*Assumptions.* The augmented matrix $\tilde{A} = A + v\ell$ is nonnegative, irreducible and viable: its Perron–Frobenius root λ satisfies $0 < \lambda < 1$, and $r^* = 1/\lambda - 1 > 0$. Irreducibility means that every commodity is basic in the augmented system; it makes λ a simple root and the left and right eigenvectors, $p^*\tilde{A} = \lambda p^*$ and $\tilde{A}q^* = \lambda q^*$, strictly positive. A commodity is basic here if it enters, directly or indirectly, the production of every commodity in the augmented system; individual input coefficients may be zero. The paper's technique satisfies irreducibility, since A > 0. Along a path the technique, the real wage and the unit values are fixed, and so are $k_m$ (S1), $s_m = \Lambda_m - k_m$, the value rates of profit $\rho_m = s_m/k_m$, q*, r* and total capital $K_T$; outputs are $Q_m = K_m/k_m$ (S2). At each state, prices satisfy equations (1)–(2) of the paper and the anchor condition $r_a = \bar{r}$, and the sectoral rates are $r_m = p_m/c_m^p - 1$ (S4).

*Shares.* Four sets of capital shares, defined in Section S1.8, must be kept apart. At current prices, $\varphi_m = c_m^p Q_m/\Sigma_n c_n^p Q_n$ are the actual shares, and $\varphi^*_m = c_m^p q^*_m/\Sigma_n c_n^p q^*_n$ those of the reference (standard) production at the same prices. In value, $\kappa_m = K_m/K_T$ are the actual shares and $\kappa^*_m = k_m q^*_m/\Sigma_n k_n q^*_n$ the reference shares. The two price-valued weightings average the same rates to different values at every state (Section S1.8): the *actual average* and the *reference average* are

$$\sum_m \varphi_m r_m = \bar{r} = \frac{S}{K_T} = \sum_m \kappa_m \rho_m, \sum_m \varphi^*_m r_m = r^*.$$

The weights $\varphi^*_m$ move with prices; q*, κ* and r* do not. On q, κ and φ, the star marks the reference system; on z, K, r and p, it marks equalization: z* and K* are equalized allocations, which are not, in general, the reference allocation, and r* and p* are the uniform rate and the production prices. The rate r* is also the rate of the reference system (Section S1.8).

*Migration map.* The anchored sector neither gives nor receives capital, because its rate always lies between the two others (step (a) of the proof of Proposition 1), so that $K_a$ stays fixed and the state moves along a segment of the simplex parallel to the side opposite vertex a. Label the two other sectors i and j once and for all, and set $z = K_i$, $K_j = K_T - K_a - z$ and $g(z) = r_i(z) - r_j(z)$. Capital flows to i when $g > 0$ and to j when $g < 0$, so that the gap (S6) between receiver and donor is $|g|$. The migration rule of the paper is the map

$$T(z) = z + \alpha \frac{g(z)}{r_d(z)},$$

where $r_d$ is the donor's rate: $r_j$ when $g > 0$, $r_i$ when $g < 0$. The rule is defined where the donor's rate is positive. At an equalized allocation z*, g(z*) = 0; all rates then equal r* and prices are proportional to p* (Sections S1.8 and S1.9.3). An equalized allocation is *regular* if the three price equations have a unique solution that depends smoothly on z near it. Primes denote derivatives with respect to z, taken at z* unless stated otherwise.

*Linear stability.* Because g(z*) = 0 and both possible donors earn r* at z*, the change of donor at z* does not affect the derivative of T: the term involving the derivative of $r_d$ is multiplied by g(z*) = 0. Hence

$$T'(z^*) = \mu = 1 + \frac{\alpha}{r^*} g'(z^*),$$

as in (S7); the fixed labeling changes nothing, since $d(r_h - r_l)/dK_h = r_i' - r_j'$ whichever of i and j receives. Thus z* is strictly attracting for the linearization, $|\mu| < 1$, if and only if (S8) holds. The economic part of (S8), $g' < 0$, says only that $r_i' < r_j'$, which two positive derivatives could satisfy. The two averages determine the signs.

*S1.9.2 The receiver's profit rate (Proposition 1)*

**Proposition 1.** Let z* be a regular interior equalized allocation with g′(z*) < 0. Then

$$r_i' < r_a' < r_j' \text{ and } r_i' < 0 < r_j'.$$

Consequently, there is an interval around z* in which every transfer between two states, from the less to the more profitable sector, lowers the receiver's profit rate and raises the donor's, whether the receiver is i (z < z*) or j (z > z*), and whatever the movement of the receiver's selling price.

*Proof.* (a) *Ordering, from the actual average and the anchor.* Since the $\varphi_m$ sum to one and $r_a = \bar{r}$, the actual average gives, at every state,

$$\varphi_i(r_i - r_a) + \varphi_j(r_j - r_a) = 0,$$

so that

$$r_a = \frac{\varphi_i r_i + \varphi_j r_j}{\varphi_i + \varphi_j}$$

lies strictly between $r_i$ and $r_j$ when they differ. Differentiating at z*, the terms in $\varphi_i'$ and $\varphi_j'$ multiply the deviations $r_i - r_a$ and $r_j - r_a$, which vanish there, so that no assumption of constant weights is needed:

$$\varphi_i(r_i' - r_a') + \varphi_j(r_j' - r_a') = 0$$

Rearranging,

$$r_a' = \frac{\varphi_i r_i' + \varphi_j r_j'}{\varphi_i + \varphi_j},$$

at every regular equalized allocation. Hence $r_a'$ is a positive weighted average of $r_i'$ and $r_j'$, and g′ < 0 gives $r_i' < r_a' < r_j'$.

(b) *Signs, from the reference average.* The reference average equals r* at every state, and r* is constant along the path. Differentiating gives

$$\sum_m \varphi_m^{*}{}' r_m + \sum_m \varphi_m^{*} r_m{}' = 0.$$

At z* the first sum equals $r^*\Sigma_m\varphi^*_m{}' = 0$, because the weights always sum to one, so that

$$\varphi_i^{*} r_i{}' + \varphi_a^{*} r_a{}' + \varphi_j^{*} r_j{}' = 0$$

at every regular equalized allocation. Three strictly ordered numbers whose average with strictly positive weights is zero have a negative smallest element and a positive largest one: $r_i' < 0 < r_j'$. The sign of $r_a'$ is not determined, and it is not needed. If the smallest derivative were nonnegative, the strictly larger derivatives would make the weighted sum positive; if the largest were nonpositive, the sum would be negative.

(c) *Transfers.* Near a regular z*, prices and rates are continuously differentiable functions of z (rational functions, as in step (a) of the proof of Proposition 2), so that the strict inequalities $r_i' < 0 < r_j'$ and g′ < 0 hold on an interval around z*. Since $g(z^*)=0$ and $g'<0$ throughout this interval, $g>0$ for $z<z^*$: sector i receives capital, z rises and $r_i$ falls. For z > z*, g < 0: sector j receives, z falls, and $r_j$, which increases with z, falls. For a transfer between two states of the interval, the change in the receiver's rate is the integral of a derivative of constant sign, and the donor's rate rises in both cases. ∎

The actual average and the anchor order the derivatives; the constancy of the reference average forces the extreme ones to have opposite signs. Since $1 + r_h = p_h/c_h^p$ (S4), the receiver's rate falls between two states, labeled (0) before the transfer and (1) after it, if and only if

$$\frac{p_h^{(1)}}{p_h^{(0)}} < \frac{c_h^{p(1)}}{c_h^{p(0)}}.$$

When its selling price rises, its unit cost, valued at current prices, therefore rises proportionally more: with an initial rate of 20 percent, a price rise of 2 percent and a unit-cost rise of 3 percent give $1 + r = 1.20 \times 1.02/1.03 \approx 1.188$. Nor does the fall in the rate require a fall in the mass of profit, which can grow, but more slowly than the capital valued at current prices (Section S1.5). Because (S4) contains no quantity, an inflow of capital at unchanged prices would leave the rate unchanged (Section S1.1): the fall operates entirely through the prices selected by the two aggregate equalities and the anchor, without any excess-supply price rule.

*S1.9.3 Regularity, the stability condition and the direction of transfers (Proposition 2)*

The line

$$\sum_m \kappa_m = 1, \sum_m \kappa_m \rho_m = r^*$$

of Section S1.8 is the set of equalized allocations. Uniform rates require $\bar{r} = r^*$ (Section S1.8). Conversely, at every interior allocation on the line (at κ* itself the price system is singular, and these prices are one solution among a continuum), prices $p = \gamma p^*$, with γ such that $p \cdot Q = W_T = (1 + r^*)K_T$, satisfy both equalities, since $p\tilde{A}Q = \lambda p \cdot Q = K_T$ and total profit is then $W_T - K_T = r^*K_T = S$; all rates equal $r^* = \bar{r}$, so that the anchor condition holds as well. The reference proportions lie on this line: since $k = \Lambda\tilde{A}$ and $\tilde{A}q^* = \lambda q^*$,

$$\sum_m k_m q_m^* = \lambda \sum_m \Lambda_m q_m^*,$$

so that the reference production yields surplus value

$$\sum_m s_m q_m^* = (1-\lambda) \sum_m \Lambda_m q_m^* \text{ and } \sum_m \kappa_m^* \rho_m = \frac{1-\lambda}{\lambda} = r^*.$$

In the simplex of capital shares, the singular line of anchor $a$ passes through vertex $a$ and $\kappa^*$. An allocation lies on it if and only if $E=0$, where

$$E = \kappa_i^* \kappa_j - \kappa_j^* \kappa_i = \frac{k_i k_j \left(q_i^* Q_j - q_j^* Q_i\right)}{K_T \sum_n k_n q_n^*}.$$

In the interior, $E=0$ is equivalent to $\kappa_j/\kappa_i = \kappa_j^*/\kappa_i^*$, or $Q_j/Q_i = q_j^*/q_i^*$: only the ratio of the two non-anchored sectors is fixed. The anchored share completes the total to one. Thus $E$ locates the singular line and its sign distinguishes its two sides.

**Proposition 2.** Let z* be an interior equalized allocation off the singular line of anchor a, with $\rho_i \neq \rho_j$. Then the price system is regular near z*, and z* is strictly locally stable for some $\alpha > 0$, and then exactly for $0 < \alpha < -2r^*/g'(z^*)$, if and only if

$$\left(\rho_a - r^*\right)\left(\kappa_a - \kappa_a^*\right) > 0 \qquad \text{(S12)}$$

More precisely, with $D = \varphi^*_i\varphi_j - \varphi^*_j\varphi_i$, which is nonzero,

$$g'(z^*) = \frac{-\rho_i - \rho_j}{K_T} \cdot \frac{\varphi_i + \varphi_j}{D} \qquad \text{(S13)}$$

and sign $g'(z^*) = -\text{sign}[(\rho_a - r^*)(\kappa_a - \kappa^*_a)]$.

*Proof.* (a) *Regularity.* The three price equations,

$$p \cdot Q = W_T, \, p\tilde{A}Q = K_T, \, p_a - (1+\bar{r})(p\tilde{A})_a = 0,$$

are linear in $p$, with coefficients that depend smoothly on $z$. At $z = z^*$ we have $\acute{r} = r^*$. Let $\eta$ be a row-vector solution of its homogeneous system, obtained by replacing $W_T$ and $K_T$ by zero. For this square system, invertibility is equivalent to the homogeneous system having only the zero solution. Set

$$Y = \eta\left[I - (1+r^*)\widetilde{A}\right].$$

Since $\left[I - (1+r^*)\widetilde{A}\right]q^* = 0$, $Y \cdot q^* = 0$. The homogeneous equations also give $\eta Q = \eta \widetilde{A}\, Q = 0$, hence $Y \cdot Q = \eta Q - (1+r^*)\eta \widetilde{A}\, Q = 0$; the homogeneous anchor equation is $Y_a = 0$.

Off the singular line, $q^*$, $Q$ and the unit vector $u_a$ are linearly independent, since their determinant is $\pm(q^*_iQ_j - q^*_jQ_i)$, so that $Y = 0$: a row vector that annihilates a basis of three-dimensional space must be zero.

The equality $Y = 0$ is equivalent to $\eta \widetilde{A} = \lambda\eta$, with $\lambda = 1/(1+r^*)$. Since $\lambda$ is simple, $\eta = c\, p^*$ for some scalar $c$. Then $0 = \eta \widetilde{A}\, Q = c\, p^* \widetilde{A}\, Q$; positivity of $p^* \widetilde{A}\, Q$ forces $c = 0$ and hence $\eta = 0$.

Invertibility at $z^*$ gives a nonzero determinant. By continuity, the determinant remains nonzero in a neighborhood of $z^*$. Cramer's rule expresses the prices there as rational functions of $z$, with no vanishing denominator. Since $p(z^*) = \gamma\, p^*$, $\gamma > 0$, continuity also preserves positive prices. The sectoral rates are therefore well-defined and differentiable in that neighborhood.

(b) *Formula (S13).* Local regularity allows us to compute $g'(z^*)$. All derivatives and price-valued capital shares below are evaluated at $z^*$. With $K_a$ and $K_T$ fixed,

$$\bar{r} = \sum_m \kappa_m \rho_m = \frac{\rho_a K_a + \rho_i z + \rho_j (K_T - K_a - z)}{K_T},$$

so that the anchor gives

$$r_a' = \bar{r}' = \frac{\rho_i - \rho_j}{K_T}.$$

Combining $g' = r_i' - r_j'$ with the identity in step (a) of the proof of Proposition 1 gives

$$r_i' = r_a' + \frac{\varphi_j g'}{\varphi_i + \varphi_j}, r_j' = r_a' - \frac{\varphi_i g'}{\varphi_i + \varphi_j}.$$

Substituting these expressions for $r_i'$ and $r_j'$ into $\sum_m \varphi_m^* r'_m = 0$, established in step (b) of the proof of Proposition 1, and collecting terms gives, since $\sum_m \varphi_m^* = 1$,

$$0 = (\varphi_a^* + \varphi_i^* + \varphi_j^*) r_a' + \frac{g'}{\varphi_i + \varphi_j} (\varphi_i^* \varphi_j - \varphi_j^* \varphi_i)$$
$$0 = r_a' + \frac{D g'}{\varphi_i + \varphi_j}.$$

At $z^*$, $\sum_n c_n^p Q_n = K_T$. Using $\varphi_m = c_m^p Q_m / K_T$ and $\varphi_m^* = c_m^p q_m^* / \sum_n c_n^p q_n^*$ in $D = \varphi_i^* \varphi_j - \varphi_j^* \varphi_i$ gives

$$D = \frac{(c_i^p q_i^*)(c_j^p Q_j) - (c_j^p q_j^*)(c_i^p Q_i)}{K_T \sum_n c_n^p q_n^*}$$
$$D = \frac{c_i^p c_j^p (q_i^* Q_j - q_j^* Q_i)}{K_T \sum_n c_n^p q_n^*}.$$

Comparison with the expression for $E$ shows that $D$ is a positive multiple of $E$, hence nonzero off the singular line. Inserting $r_a' = (\rho_i - \rho_j)/K_T$ into the equation above and solving for $g'$ therefore gives (S13):

$$\frac{\rho_i - \rho_j}{K_T} + \frac{D g'}{\varphi_i + \varphi_j} = 0$$
$$\Rightarrow g'(z^*) = \frac{-\rho_i - \rho_j}{K_T} \frac{\varphi_i + \varphi_j}{D}.$$

(c) *Sign.* Write κ = κ* + Δκ. Since κ and κ* both lie on the line of equalized allocations,

$$\Delta\kappa_i + \Delta\kappa_j = -\Delta\kappa_a, \rho_i \Delta\kappa_i + \rho_j \Delta\kappa_j = -\rho_a \Delta\kappa_a,$$

whence

$$\Delta\kappa_i = \frac{-(\rho_a - \rho_j)\Delta\kappa_a}{\rho_i - \rho_j}, \Delta\kappa_j = \frac{(\rho_a - \rho_i)\Delta\kappa_a}{\rho_i - \rho_j}.$$

Substituting these deviations into $E = \kappa_i^* \Delta\kappa_j - \kappa_j^* \Delta\kappa_i$ gives

$$E = \frac{\left[\kappa_i^*(\rho_a - \rho_i) + \kappa_j^*(\rho_a - \rho_j)\right]\Delta\kappa_a}{\rho_i - \rho_j}.$$

The bracket equals $\rho_a - r^*$, since $\kappa_i^* + \kappa_j^* = 1 - \kappa_a^*$ and $\kappa_i^* \rho_i + \kappa_j^* \rho_j = r^* - \kappa_a^* \rho_a$. Thus

$$E = \frac{(\rho_a - r^*)(\kappa_a - \kappa_a^*)}{\rho_i - \rho_j}.$$

As D has the sign of E,

$$\operatorname{sign} g' = -\operatorname{sign} \frac{\rho_i - \rho_j}{E} = -\operatorname{sign}\left[(\rho_a - r^*)(\kappa_a - \kappa_a^*)\right],$$

because $(\rho_i - \rho_j)^2 > 0$. Strict stability requires $g' < 0$, which is (S12), and $\alpha < -2r^*/g'$, which holds for small enough α because g′ is finite. ∎

*Reading.* Condition (S12) involves only technical and value data: the anchored sector's value rate of profit relative to r*, and its capital share relative to its reference share. Since $\kappa_a$ is fixed along a path, the condition can be checked at the starting allocation. If $\rho_i = \rho_j \neq \rho_a$, every equalized allocation has $\kappa_a = \kappa^*_a$, so that (S12) fails, and $\bar{r}$ is constant along the segment, so that $g' = 0$ on its regular portions: there is no strict attraction of an isolated equalized allocation. On the singular line the capital differs from the reference proportions only through $\kappa_a$, so that $\bar{r} - r^* = (\rho_a - r^*)(\kappa_a - \kappa^*_a)/(1 - \kappa^*_a)$ there: (S12) states that, where the segment of anchor a crosses the singular line, the surplus value produced exceeds $r^*K_T$, that is, $\Gamma < 0$.

*Direction of transfers along the whole segment.* Condition (S12) also governs the direction of every transfer on the segment, not only near z*. By (S11), $r_i - \bar{r}$ has the sign of $\Gamma/(\sigma_i - \sigma_j)$, and so does g, since $r_a = \bar{r}$ lies between $r_i$ and $r_j$. At every state of the segment with positive prices, $\sigma_i - \sigma_j$ has the sign of $q^*_iQ_j - q^*_jQ_i$, an affine function of z that vanishes only at the point $z_s$ where the segment crosses the singular line, and $\Gamma = r^*K_T - S$ is an affine function of z that vanishes only at z*. Hence g changes sign only at z* and at $z_s$. On the side of $z_s$ that contains z*, g has the sign of $g'(z^*)(z - z^*)$; beyond $z_s$, its sign is the opposite of its sign between $z_s$ and z*. A transfer moves z in the direction given by the sign of g when the donor's rate is positive. Hence, at every state with positive prices and profit rates: when (S12) holds, every transfer from a state on the same side of the singular line as z* is directed toward z*, and every transfer from a state beyond that line is directed away from it; when (S12) fails, every transfer from a state on the side of z* is directed away from z*. Convergence to z* therefore requires a starting allocation on the same side of the singular line as z*. On that side the direction of capital movement is always the right one; what can still prevent convergence is the size of the steps or the loss of positive prices or profit rates.

*A stabilizing anchor exists.* Since κ and κ* both lie on the line of equalized allocations,

$$\sum_m \left(\kappa_m-\kappa_m^*\right)\left(\rho_m-r^*\right)=\sum_m \kappa_m \rho_m-\sum_m \kappa_m^* \rho_m-r^*\left(\sum_m \kappa_m-\sum_m \kappa_m^*\right)=0.$$

The three terms of this sum are the left-hand sides of (S12) for the three possible anchors. At any equalized allocation other than κ*, provided the value rates are not all equal, the three terms cannot all vanish; their zero sum therefore implies that at least one is positive and at least one negative: some anchor makes the allocation locally stable and some other anchor makes it unstable. At the bread-anchored equilibrium (0.5600, 0.2514, 0.1886), the terms are +0.0003 for bread, −0.0015 for iron and +0.0013 for meat, and g′(z*) is negative under the bread and meat anchors and positive under the iron anchor. Stability is therefore a property of the rule that selects prices within the admissible set, not of the two equalities, and a stabilizing selection of this kind exists at every interior equalized allocation other than the reference point, when the value rates are not all equal. A positive term also gives ρi ≠ ρj and E ≠ 0 for that anchor, since ρi = ρj would force κa = κ*a, so that Proposition 2 applies to it. A demand-side mechanism such as a response of prices to excess supply acts on the same degree of freedom and could select a stabilizing path; it is not modeled here.

### *S1.9.4 Existence of convergent paths (Proposition 3)*

Propositions 1 and 2 describe equalized allocations that would correct deviations if they were reached. That paths actually reach them must be proved separately.

**Proposition 3.** Let a be an anchor with $\rho_a \neq r^*$ such that the two other sectors have $\rho_i \neq \rho_j$.

(i) Every interior equalized allocation z* that satisfies (S12) is regular, with g′(z*) < 0, and for 0 < α < −2r*/g′(z*) it attracts an interval J of starting allocations on its segment: from every $z_0$ in J, the migration rule converges to z*, all rates converge to r*, and Proposition 1 applies at every transfer. If α < −r*/g′(z*), the approach is monotone, without overshooting.

(ii) These allocations form the part of the line of equalized allocations, inside the simplex, that lies on the side of κ* given by (S12); those close enough to κ* also have a strictly positive net product (I − Ã)Q, so that the states along the paths of (i) that reach them are reproducible. If the $\rho_m$ are not all equal, an anchor of this kind exists. Every viable three-sector technique in which every commodity is basic and the value rates of profit are not all equal therefore has convergent paths of this kind, through reproducible states.

*Proof.* (a) *Regularity and sign.* By step (c) of the proof of Proposition 2, E = $(\rho_a - r^*)(\kappa_a - \kappa^*_a)/(\rho_i - \rho_j)$ at every equalized allocation. Under (S12), E ≠ 0, so that z* lies off the singular line, is regular and has g′(z*) < 0 by Proposition 2.

(b) *Contraction.* Near z*, $r_i$ and $r_j$ are close to r* > 0, and on each side of z*, T coincides with z + $\alpha g/r_j$ or z + $\alpha g/r_i$, functions that are continuously differentiable on a whole neighborhood of z*. Their derivatives at z* both equal μ (Section S1.9.1), so that T is continuously differentiable near z*, with T(z*) = z* and T′(z*) = μ. The bound on α gives |μ| < 1. Fix L with |μ| < L < 1, and choose δ > 0 such that |T′(z)| ≤ L on J = [z* − δ, z* + δ] and such that on J capitals and prices are positive, the price system is regular, the net product remains positive if it is positive at z*, and the conclusions of Proposition 1 hold. By the mean value theorem, for z in J and some z̃ between z and z*,

$$\left|T(z)-z^*\right|=\left|T'(\tilde{z})\right|\left|z-z^*\right|\le L\left|z-z^*\right|.$$

Since $\left|z-z^*\right|\le\delta$ on $J$, the bound gives $\left|T(z)-z^*\right|\le L\delta<\delta$: T maps J into itself. Iterating the same bound from any $z_0\in J$ gives

$$\left|z_t-z^*\right|\le L^t\left|z_0-z^*\right|,$$

which tends to zero. By continuity, prices and rates converge to their values at z*, where all rates equal r*. Every transfer joins two states of J, so that Proposition 1 applies to it.

(c) *Monotone approach.* If $\alpha < -r^*/g'(z^*)$, then $0 < \mu < 1$, and δ can be chosen so that $T' > 0$ on J. Then $T(z) - z^*$ has the sign of $z - z^*$: the path approaches $z^*$ from one side, and the receiving sector does not change. Together with the contraction bound, preservation of the side of $z^*$ gives monotone convergence.

(d) *Existence.* The point $\kappa^*$ lies on the line of equalized allocations (Section S1.9.3) and has positive coordinates. Along the line, set $\kappa_a = \kappa^*_a + s$; the two equalities that define the line then fix $\kappa_i$ and $\kappa_j$ as in step (c) of the proof of Proposition 2, and all three shares remain positive for small |s|. Choosing a small nonzero s with the sign of $\rho_a - r^*$ therefore gives (S12), so that the interior allocations satisfying (S12) are the points of the line on that side of $\kappa^*$, up to the boundary of the simplex. At each of them, the prices $\gamma p^*$ of Section S1.9.3 give uniform rates. Near $\kappa^*$, the net product $(I - \tilde{A})Q$ is also strictly positive, as it is at $\kappa^*$, where Q is proportional to $q^*$ and $(I - \tilde{A})q^* = (1 - \lambda)q^*$. Finally, if the $\rho_m$ are not all equal, $\min_m\rho_m < r^* < \max_m\rho_m$, since $r^* = \Sigma_m\kappa^*_m\rho_m$ with positive weights. If the three $\rho_m$ are distinct, at most one equals $r^*$, and any other sector can serve as anchor; the two remaining sectors have distinct value rates. If two of them are equal, take one of these two sectors as anchor: its value rate differs from $r^*$, which lies strictly between the two distinct values. ∎

Steps (a)–(c) use only regularity and $g'(z^*) < 0$. As far as the price system and the migration rule are concerned, Proposition 3(i) therefore applies to every equalized allocation that satisfies (S12), not only to those near $\kappa^*$, and step (d) shows that such allocations exist for every technique. Reproducibility, a nonnegative net product, is a separate condition on the allocation: it holds near $\kappa^*$ but not everywhere on the line. The bound on α depends on the allocation: as $z^*$ approaches $\kappa^*$, which lies on the singular line, D tends to zero in (S13), $g'(z^*)$ diverges and the admissible α shrinks, so that no single α serves every equalized allocation. Proposition 3 is an existence result. It does not assert that every anchor, or every α fixed in advance, yields convergence, and it does not bound the basin of attraction beyond J; Section S1.9.3 shows only that the basin lies on the same side of the singular line as z*. For the paper's technique, Figure S1 shows that the basin covers nearly all of that side.

Taken together, Propositions 1–3 establish the following for every viable three-sector technique in which every commodity is basic and the value rates of profit are not all equal. There are anchors, anchored capitals and migration coefficients for which a continuum of non-uniform, reproducible starting allocations leads to uniform profitability. Along these paths, capital moves from the less to the more profitable sector, the receiver's output rises, its profit rate falls and the donor's rises at every transfer, and all rates converge to r*. No assumption is made about the direction of the receiver's selling price, and no demand function is involved.

*S1.9.5 Relative prices (Theorem)*

Propositions 1–3 concern profit rates; the theorem concerns prices. Let

$$\Psi_{mn} = \frac{p_n \tilde{a}_{nm}}{c_m^p}$$

be the share of input n, wage goods included, in the unit cost of m. Each row sums to one because $\sum_n p_n \tilde{a}_{nm} = c_m^p$. Its positive entries follow the input links of $\tilde{A}$ in reverse orientation, so Ψ is also irreducible.

At fixed technique,

$$d \ln c_m^p = \frac{\sum_n \tilde{a}_{nm} d p_n}{c_m^p} = \sum_n \Psi_{mn} d \ln p_n .$$

Since 1 + $r_m = p_m/c_m^p$, at every state

$$d\ln(1+r_m) = d\ln p_m - \sum_n \Psi_{mn} d\ln p_n:$$

the proportional change in its gross return factor $1+r_m$ is the proportional change in its price minus a weighted average, with weights $\Psi_{mn}$, of the proportional changes in the prices of its inputs. At $z^*$, where all rates equal $r^*$, set $\xi_m = d\ln p_m/dz$ and let $\xi = (\xi_m)_m$ and $r' = (r'_m)_m$ be column vectors of log-price and profit-rate derivatives. The relation becomes

$$\frac{r_m'}{1+r^*} = \xi_m - \sum_n \Psi_{mn}\xi_n \text{ for every } m,$$

or, in vector form,
$$(I-\Psi)\xi = \frac{r'}{1+r^*}.$$

Row-stochasticity gives $(I-\Psi)1 = 0$, where **1** is the vector of ones. To show that the kernel contains only scalar multiples of 1, let $h = \Psi h$ and choose $h_m = H = max_n h_n$. Then $0 = \sum_n \Psi_{mn}(H-h_n)$. Every summand is nonnegative, so $\Psi_{mn} > 0$ forces $h_n = H$. Every positive input link from a maximal component therefore leads to a maximal component. Repeating this argument along a path and using irreducibility reaches every commodity, so $h = H1$. Thus $ker(I-\Psi) = span\{1\}$.

This kernel describes precisely what the differentiated price equations leave undetermined. If $\xi$ is a solution, adding the same scalar $\zeta$ to every component leaves its right-hand side unchanged:

$$(I-\Psi)(\xi+\zeta 1) = (I-\Psi)\xi + \zeta(I-\Psi)1 = \frac{r'}{1+r^*}.$$

Conversely, any two solutions $\xi$ and $\xi^{(0)}$ satisfy $(I-\Psi)(\xi-\xi^{(0)})=0$, so

$$\xi=\xi^{(0)}+\zeta 1, \xi_i-\xi_j=\xi_i^{(0)}-\xi_j^{(0)}.$$

Thus one particular solution determines every derivative of a relative price. The common proportional price movement per unit of $z$, represented by $\zeta$, is fixed by the aggregate constraint below.

**Theorem. Let z* be an interior equalized allocation off** the singular line of anchor a. (i) The derivative d $\ln(p_i/p_j)$/dz has the sign of g′(z*): on either side of z* and near it, the receiver's price relative to the donor's falls, to first order, if and only if g′(z*) < 0, that is, under (S12) when $\rho_i \neq \rho_j$. (ii) Its size is given by (S14) below.

*Proof of (i).* Let $\nu = -g'/(\varphi_i + \varphi_j)$. Step (b) of the proof of Proposition 2 gives $r_i' = r_a' - \nu\varphi_j$, $r_j' = r_a' + \nu\varphi_i$ and $r_a' = \nu D$, with $D = \varphi^*_i\varphi_j - \varphi^*_j\varphi_i$. Substituting D and collecting terms gives

$$r_i'=-\nu\left[\varphi_j\left(1-\varphi_i^*\right)+\varphi_i\varphi_j^*\right], r_j'=\nu\left[\varphi_i\left(1-\varphi_j^*\right)+\varphi_j\varphi_i^*\right].$$

Both brackets are positive, so that $r_i'$ has the sign of g′ and $r_j'$ the opposite sign, as in Proposition 1 when g′ < 0.

Suppose g′ < 0, so that $r_i' < 0 < r_j'$, and suppose, to the contrary, that $\xi_i \geq \xi_j$. Since $r_i' < 0$, $\xi_i$ is smaller than the average of the $\xi_n$ of its inputs, with weights $\Psi_{in}$, so that some other commodity has $\xi_n > \xi_i$; as $\xi_j \leq \xi_i$, it can only be a, so that $\xi_a > \xi_i$.

Since $r_j' > 0$, $\xi_j$ exceeds the average of the $\xi_n$ of its inputs, so that some other commodity has $\xi_n < \xi_j$; as $\xi_i \geq \xi_j$, it can only be a, so that $\xi_a < \xi_j$. Then

$$\xi_a>\xi_i\geq\xi_j>\xi_a,$$

which is impossible. Hence $\xi_i < \xi_j$, that is, d $\ln(p_i/p_j)$/dz < 0.

If $g' > 0$, the same argument with i and j exchanged gives $\xi_i > \xi_j$. If $g' = 0$, then $r' = 0$, and since the kernel of $I - \Psi$ is spanned by **1**, $\Psi$ being irreducible, all $\xi_m$ are equal and the relative price is stationary to first order. For $g' \neq 0$, continuity preserves this strict sign near $z^*$; since i receives capital when $z < z^*$ and j when $z > z^*$, the receiver's relative price falls on either side when $g' < 0$. ∎

The argument uses only the signs of the rate changes and the fact that a change in unit cost is an average of input-price changes. The receiver, whose rate falls, has a price that rises less than the average of its input prices; the donor, whose rate rises, has a price that rises more; with three commodities, the third cannot lie both above the first and below the second. Part (ii) gives the size of the response.

Part (ii) of the theorem states that

$$\left(1+r^*\right)\frac{d\ln\left(p_i/p_j\right)}{dz}=g'\,\frac{\varphi_i^*\varphi_j\tau_{ji}+\varphi_j^*\varphi_i\tau_{ij}}{\varphi_i+\varphi_j} \tag{S14}$$

The fraction is positive, in agreement with (i), and gives the response of the relative price per unit of $g'$.

*Proof of (ii). Step 1 (rates).* By the proof of (i), $r_i' = r_a' - \nu\varphi_j$, $r_j' = r_a' + \nu\varphi_i$ and $r_a' = \nu(\varphi_i^*\varphi_j - \varphi_j^*\varphi_i)$, that is,

$$r'=\nu\left[\varphi_j\left(\varphi_i^*1-u_i\right)-\varphi_i\left(\varphi_j^*1-u_j\right)\right],$$

where $u_m$ is the m-th unit vector: the rate-change vector is the difference of two profiles, each a uniform change outside one sector compensated at that sector.

*Step 2 (first-passage times).* For a target n, set $\tau_{nn} = 0$ and, for $m \neq n$, let $\tau_{mn}$ be the mean number of input links followed from m until the first encounter with n, in the chain whose transition probabilities are the cost shares Ψ. Conditioning on the first link gives

$$\tau_{mn} = 1 + \sum_k \Psi_{mk} \tau_{kn}, m \neq n.$$

The unit term counts that link, the sum the links still needed. The solution exists and is positive because deleting row and column n from the irreducible stochastic matrix Ψ leaves a substochastic matrix with spectral radius below one. Let $\tau^{(n)} = (\tau_{mn})_m$. By the recurrence, $\left[(I - \Psi)\tau^{(n)}\right]_m = 1$ for $m \neq n$. At z*, $c^p = \lambda p$, so that φ* is the stationary vector of Ψ:

$$(\varphi^* \Psi)_n = \frac{p_n (\widetilde{A} q^*)_n}{\sum_m c^p_m q^*_m} = \frac{\lambda p_n q^*_n}{\sum_m c^p_m q^*_m} = \varphi^*_n,$$

Hence $\varphi^*(I - \Psi)\tau^{(n)} = 0$, which fixes the n-th component: $\left[(I - \Psi)\tau^{(n)}\right]_n = 1 - 1/\varphi^*_n$. Multiplying by $\varphi_n^*$ gives

$$(I - \Psi)(\varphi^*_n \tau^{(n)}) = \varphi^*_n \mathbf{1} - u_n.$$

*Step 3 (conclusion).* By Steps 1 and 2 and the linearity of I − Ψ,

$$\xi^{(0)} = \frac{v}{1 + r^*} \left[\varphi_j \varphi^*_i \tau^{(i)} - \varphi_i \varphi^*_j \tau^{(j)}\right]$$

solves $(I - \Psi)\xi = r'/(1 + r^*)$. Every solution is $\xi = \xi^{(0)} + \zeta\mathbf{1}$ for some scalar ζ, which cancels from $\xi_i - \xi_j$ and is fixed by the price level (consequence 2 below). Since $\tau_{ii} = \tau_{jj} = 0$,

$$(1 + r^*)(\xi_i - \xi_j) = -v\left(\varphi_j \varphi^*_i \tau_{ji} + \varphi_i \varphi^*_j \tau_{ij}\right),$$

and substituting $-\nu = g'/(\varphi_i + \varphi_j)$ gives (S14). ∎

*Interpretation.* Differentiating $\ln p_m = \ln(1 + r_m) + \ln c_m^p$ shows that a change in a sector's gross return is transmitted to its price, and from there to the prices of the sectors that use it as an input, with the cost shares as weights; replacing an input price by its own price–cost expression adds a further profit layer, as in Sraffa's (1960, ch. VI) reduction to dated quantities of labor. Followed upstream from m, the chains of inputs stop at their first encounter with n, since everything beyond it is already contained in $p_n$, to which $p_m$ is compared; $\tau_{mn}$ is the expected number of profit layers met before that encounter, and the two terms of (S14) are the responses to the two profiles of Step 1. The chain is a device for tracing input-cost dependence, not an assumption about firm behavior, and forming $p_m/p_n$ fixes no numéraire.

Two consequences follow at a stable equalized allocation ($g' < 0$).

1. *The receiver's absolute price.* At $z^*$, $\xi_i<\xi_j$, so at least one of $\xi_i<0$ and $\xi_j>0$ holds; otherwise $\xi_i \geq 0 \geq \xi_j$, a contradiction. Since prices are positive, these inequalities are equivalent to $p_i'<0$ and $p_j'>0$, respectively. Hence the receiver's absolute price falls on at least one side of $z^*$. If $p_i'<0$, its price falls when i receives capital and z rises below $z^*$; if $p_j'>0$, its price falls when j receives capital and z falls above $z^*$.

Continuity preserves the relevant strict price derivative near $z^*$. Step (c) of the proof of Proposition 3 shows that, under the stricter bound on $\alpha$, the path stays on its starting side of the equilibrium. Restricting the interval J of Proposition 3, with $\alpha < -r^*/g'(z^*)$, to that side, and shortening it if necessary, gives convergent paths along which the receiver's output rises while its price and its profit rate fall at every transfer.

2. *The price level.* The equality $\Sigma_m c_m^p Q_m = K_T$, implied by (1)–(2), fixes the common movement of prices. Since $Q_i{}'=1/k_i$, $Q_j{}'=-1/k_j$ and $Q_a{}'=0$, differentiating this equality gives

$$0=\sum_m Q_m \left(c_m^p\right)'+\sum_m c_m^p Q_m{}'.$$

The second sum is $c_i^p/k_i - c_j^p/k_j$. In the first sum, use $c_m^p=\sum_n p_n \tilde{a}_{nm}$ and regroup by input:

$$\sum_m Q_m \left(c_m^p\right)'=\sum_n \left(\sum_m \tilde{a}_{nm} Q_m\right) p_n{}'.$$

Since $p_n{}'=p_n \xi_n$, this is

$$\sum_m Q_m \left(c_m^p\right)'=\sum_n p_n \left(\widetilde{A} Q\right)_n \xi_n.$$

Here $\left(\widetilde{A} Q\right)_n$ is the total quantity of commodity *n* used as an input across all sectors, wage goods included. Combining the two sums gives

$$0=\sum_m p_m \left(\widetilde{A} Q\right)_m \xi_m+\frac{c_i^p}{k_i}-\frac{c_j^p}{k_j}.$$

Dividing by $K_T$ yields

$$\sum_m \omega_m \frac{d \ln p_m}{dz}=\frac{-c_i^p/k_i-c_j^p/k_j}{K_T}, \text{with}\, \omega_m=\frac{p_m \left(\tilde{A}Q\right)_m}{K_T}.$$

The weights are positive and sum to one, since $p \widetilde{A} Q=K_T$. Each weight measures the share of total expenditure devoted to a commodity as an input across all sectors. Writing its index as *n* makes the relation to $\Psi_{mn}$ explicit:

$$\omega_n=\sum_m \frac{c_m^p Q_m}{K_T} \Psi_{mn}=\sum_m \varphi_m \Psi_{mn}.$$

The weights $\varphi_m$ are the actual sectoral capital shares: $\Psi_{mn}$ measures an input share within sector $m$, whereas $\omega_n$ aggregates that input's expenditure across sectors.

At $z^*$, substituting $\xi_m = \xi_m^{(0)} + \zeta$ into the weighted price equation and using $\sum_m \omega_m = 1$ gives

$$\sum_m \omega_m \xi_m = \sum_m \omega_m \xi_m^{(0)} + \zeta .$$

Consequently,

$$\zeta = \frac{-c_i^p / k_i - c_j^p / k_j}{K_T} - \sum_m \omega_m \xi_m^{(0)} .$$

The aggregate constraint therefore determines the common movement without choosing a commodity as numéraire.

When this level effect outweighs the relative movement, both prices move in the same direction, and on one side of z* the receiver's price rises while its unit cost rises faster; by Proposition 1, its rate still falls.

*S1.9.6 Extension to n sectors*

The two averages, the price–cost relation $(I - \Psi)\xi = r'/(1 + r^*)$ and the first-passage identity hold for any number of sectors. Suppose that every sector other than i and j is anchored at the average rate, with fixed capital. These sectors' rates then move together, with $r_m' = \bar{r}'$ for $m \neq i, j$, and lie between $r_i$ and $r_j$, so that capital moves only between i and j.

To see why the same rate calculation applies, collect the anchored sectors in the two averages. Since both sets of weights sum to one,

$$(\varphi_i + \varphi_j)\bar{r} = \varphi_i r_i + \varphi_j r_j ,$$

$$r^* - \bar{r} = \varphi_i^* (r_i - \bar{r}) + \varphi_j^* (r_j - \bar{r}) .$$

The first equation places the average between the two mobile sectors' rates. At equalization, derivatives of the weights contribute zero because every rate is $r^*$ and $\sum_m \varphi_m{}' = \sum_m \left(\varphi_m^*\right)' = 0$. Differentiation therefore yields

$$\varphi_i\left(r_i{}' - \bar{r}{}'\right) + \varphi_j\left(r_j{}' - \bar{r}{}'\right) = 0,$$

$$\varphi_i^*\left(r_i{}' - \bar{r}{}'\right) + \varphi_j^*\left(r_j{}' - \bar{r}{}'\right) = -\bar{r}{}'.$$

Together with $g' = r_i{}' - r_j{}'$, these give $r_i{}' = \bar{r}{}' - v\,\varphi_j$, $r_j{}' = \bar{r}{}' + v\,\varphi_i$ and $\bar{r}{}' = vD$, with the same $v$ and $D$ as above. Hence the rate vector used in the price proof has the same form, with every other component equal to $\bar{r}{}'$. The first-passage identity and the kernel argument also hold in any finite dimension, so the same particular solution gives (S14).

At an equalized allocation with $D \neq 0$, the regularity argument carries over. The homogeneous anchor equations make Y zero outside $\{i, j\}$; the identities $YQ = Y\,q^* = 0$ then force its two remaining components to vanish, because $q_i^* Q_j - q_j^* Q_i \neq 0$. Thus $Y = 0$. The simplicity of $\lambda$ makes the homogeneous price vector proportional to $p^*$, and the aggregate equation forces that vector to vanish, as in step (a) of the proof of Proposition 2.

The proofs of Proposition 1, of step (b) of the proof of Proposition 2 and of part (ii) of the theorem go through with the anchored sectors taken together: Proposition 1, (S13) and (S14) hold unchanged, and (S14) gives part (i) of the theorem.

Condition (S12) and Proposition 3 rely on the three-sector geometry and are not claimed for n > 3. Under other closures the rates of the remaining sectors also move, and the sign of the relative price is no longer determined by g′ alone.

*S1.9.7 Illustration: the paper's technique*

The paper's technique satisfies the assumptions (A > 0). Its value rates of profit, $(\rho_1, \rho_2, \rho_3)$ = (0.264, 0.112, 0.146), are distinct and all differ from r* = 0.2033, so that any sector can serve as anchor in Proposition 3. The two convergent Phase-0 paths illustrate Propositions 1–3. At the bread- and meat-anchored equilibria, g′(z*) < 0 (Section S1.4) and (S12) holds (Table S6). With α = 10, μ = 0.918 and 0.475, both between 0 and 1, which is the regime of step (c) of the proof of Proposition 3. The observed paths indeed approach z* monotonically from $K_{init}$, the receiving sector never changes, and every transfer lowered the receiver's rate and raised the donor's (236 of 236 and 28 of 28 transfers, Section S1.5). At the two equilibria the fraction in (S14) equals 0.986 (bread anchor) and 1.259 (meat anchor), and the receiver's price falls on both sides.

In Phase 0, since $\rho_a - r^* = e[1/(1 + OC_a) - 1/(1 + OC^*)]$, condition (S12) reads

$$\left(\mathrm{OC}^{*}-\mathrm{OC}_{a}\right)\left(\kappa_{a}-\kappa_{a}^{*}\right)>0:$$

an anchored sector of lower organic composition than the standard system ($OC_a < OC^*$ = 4.528) must be over-represented in the social capital relative to that system, and one of higher composition under-represented. At $K_{init}$, bread (OC = 3.264) is over-represented and meat (6.714) under-represented, as required, whereas iron (9.012 > OC*) is also over-represented, violating (S12).

Table S6 applies the condition at $K_{init}$; since $\kappa_a$ is fixed along each path, the condition holds at the equalized allocation if and only if it holds at the start.

**Table S6 Stability condition (S12) at $K_{init}$ and outcome of the migration rule**

| Anchor | $\rho_a - r^*$ | $\kappa_a - \kappa^*_a$ | Condition (S12) | Migration rule from $K_{init}$ |
|---|---|---|---|---|
| Bread | +0.0603 | +0.0048 | satisfied | converges (236 transfers) |
| Iron | −0.0910 | +0.0255 | violated | diverges ($r_1 < 0$ at iteration 10) |
| Meat | −0.0576 | −0.0303 | satisfied | converges (28 transfers) |

Source: Authors' simulations.

The condition is also consistent with the behavior of adjustment rules under which $\kappa_a$ changes. Under the new-capital rule, which adds $10(r_{max} - r_{min})/r_{min}$ units of new capital to the most profitable sector at each step, the anchored share falls. Under the bread anchor it falls below $\kappa^*_1 = 0.5552$ after 10 steps, and the adjustment diverges at step 17. Under the iron anchor it falls below $\kappa^*_2 = 0.2345$ after 16 steps, after which the adjustment converges, at step 44, to (0.5539, 0.2297, 0.2164). Under the meat anchor it never crosses $\kappa^*_3 = 0.2103$, and the adjustment converges at step 85. Stability thus depends jointly on the anchored sector's value rate of profit and on its capital share relative to the standard system: the iron anchor, divergent from $K_{init}$, is locally stable, for small enough $\alpha$, wherever $\kappa_2 < \kappa^*_2$.

*Any change in capital.* The proof of Proposition 1 uses only the anchor and the two averages, which hold at every state however capital changes, and r* stays constant as long as the technique and the real wage are fixed. At an equalized allocation, any small change in the capital vector, whether a transfer, new capital, a withdrawal or a combination of these, therefore moves the two non-anchored rates in opposite directions, with the anchored rate between them: no reallocation or accumulation can raise or lower all rates together. If the change lowers the receiver's rate relative to the other non-anchored rate, the receiver's rate falls, whatever happens to its price. Which changes do so depends on how they move S relative to $r^*K_T$. Since $\Gamma = 0$ at an equalized allocation, (S11) gives

$$d\left(r_h - r_l\right) = \frac{d\Gamma}{\sigma_h - \sigma_l}\left(\frac{1}{C_h^p} + \frac{1}{C_l^p}\right),$$

with $d\Gamma = (\rho_l - \rho_h)dK$ for a transfer from l to h and $d\Gamma = (r^* - \rho_h)dK$ for new capital placed in h. A transfer lowers the receiver's relative rate if $(\rho_h - \rho_l)(\sigma_h - \sigma_l) > 0$, which is the sign condition $g' < 0$ given by (S13); new capital does so if $(\rho_h - r^*)(\sigma_h - \sigma_l) > 0$. For new capital, the receiver's value rate of profit is thus compared with the uniform rate rather than with the donor's. At the bread-anchored equilibrium, new capital placed in iron lowers its rate relative to meat's, whereas new capital placed in meat raises meat's rate relative to iron's; at $K_{init}$, Table S5 likewise shows that new capital placed in meat widens Γ and raises the receiver's premium.

Figure S1 maps the outcome of the migration rule ($\alpha = 0.5$) over the simplex for the present technique. No path converged from starting points at which (S12) fails or which lie on the other side of the singular line; where both conditions hold, paths converged, or were still converging after 4,000 steps, except within 0.022 of the simplex edge, the singular line or the boundary $\kappa_a = \kappa^*_a$, where the discrete step matters. This is the pattern derived in Section

S1.9.3. For α below its bound, Proposition 3 guarantees convergence from an interval around each equalized allocation that satisfies (S12); for this technique the basin is much larger.

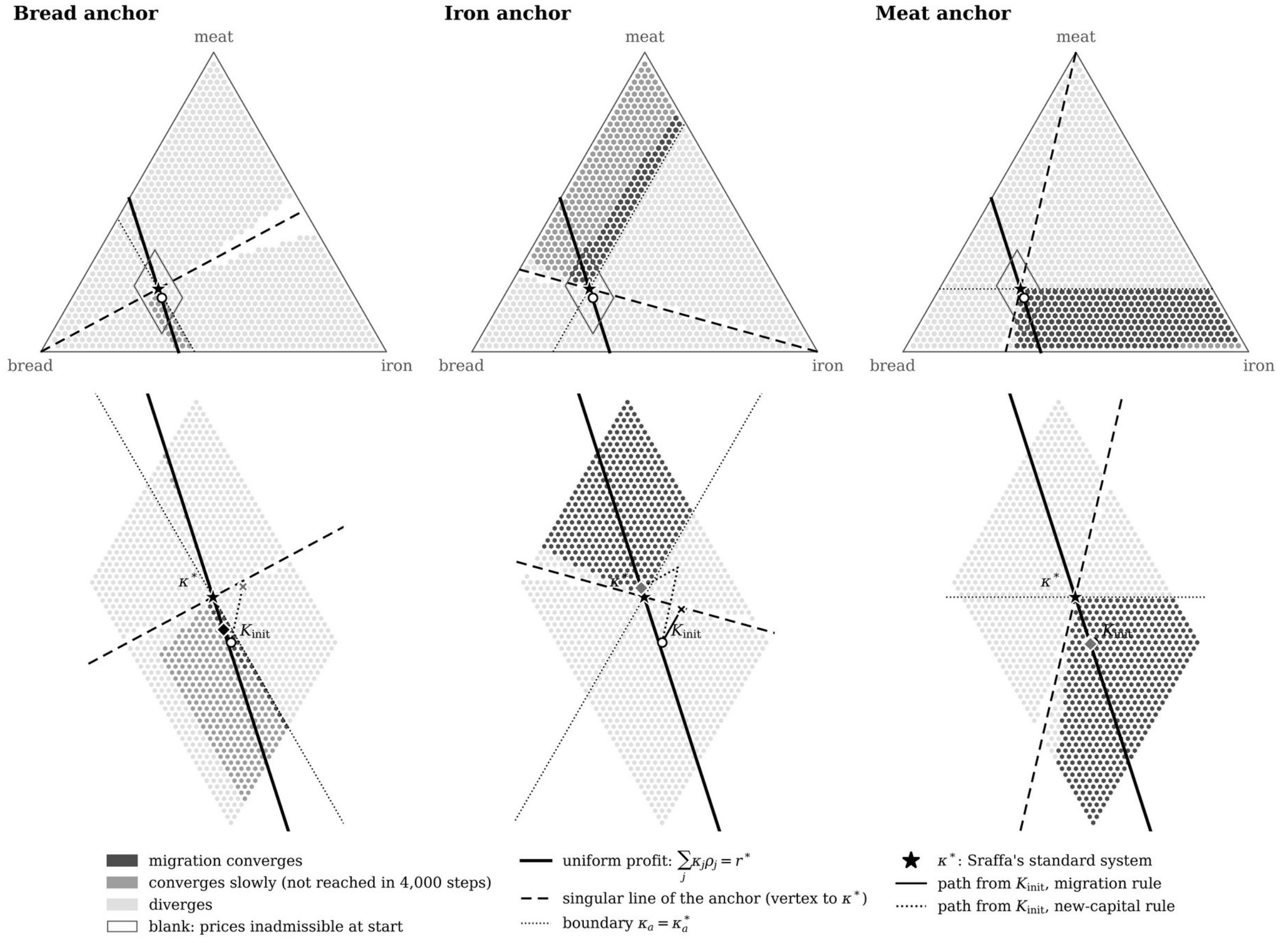

Figure S1 Outcome of the migration rule (α = 0.5, at most 4,000 steps) from every admissible starting allocation, by anchor. Top: whole simplex of capital shares (grid step 0.02). Bottom: enlargement around the standard proportions κ* (grid step 0.004), with the paths from $K_{init}$ (α = 10) under the migration rule and under the new-capital rule; diamonds mark convergence, crosses divergence. Source: Authors' simulations.

*S1.9.8 Scope and numerical checks*

*What is proved.* Identity (S10) holds for any number of sectors and any rule selecting prices within the admissible set; formula (S11) holds for three sectors under the anchor closure, for any viable technique. Propositions 1–3, the direction of transfers, the existence of a stabilizing anchor, the extension of Proposition 1 to any change in capital and the theorem hold for any viable three-sector technique in which every commodity is basic, under the anchor closure and the migration rule of the paper; Proposition 1, (S13) and (S14) extend to n sectors as stated in Section S1.9.6. Apart from the direction of transfers, these results are local: they concern equalized allocations and their neighborhoods. The simulations of the paper illustrate them and are not their basis.

*Numerical checks of the analytical results.* (S12), (S13) and (S14) were checked against complex-step derivatives, exact to machine precision, at 886,107 equalized allocations of 12,901 randomly drawn techniques with $\tilde{A}$ irreducible. Besides a baseline family ($a_{ij}$ uniform on [0, 0.35], $\ell_j$ on [0.02, 0.08], $v_j$ on [0, 2.5]), the sample includes sparse matrices, wage baskets with zero entries, r* between 0.012 and 0.05 and nearly decomposable matrices. (S12) gives the sign of g′ everywhere; the maximum relative error is $2 \times 10^{-5}$ for (S13), reached in techniques with small r*, and $1.5 \times 10^{-7}$ for (S14), whose sign is correct everywhere. At the 425,258 stable allocations, the receiver's rate falls and the donor's rises on both sides without exception; the receiver's price falls on at least one side, and on both sides in 98.7 percent of them (99.7 percent for the baseline family). For Proposition 3 and the direction of transfers, 8,398 equalized allocations were drawn along the whole line of equalized allocations of 593 baseline techniques, on both sides of κ*. The sign of g′ was as predicted at all of them, and on a grid of their segments the sign of g matched Section S1.9.3 at all 819,813 states with positive prices. With μ = 1/2, the migration rule converged monotonically, with the receiver's rate falling at every transfer, from both sides of each of the 4,223 allocations satisfying (S12): 8,445 of the 8,446 paths converged from 10 percent of the distance between z* and the singular line or the nearer end of the segment, and the last from a closer start. At 2,586 equalized allocations of 980 baseline techniques, the identity of Section S1.9.3 holds to $10^{-10}$, the sign of g′ under each of the three anchors is the opposite of its (S12) term, and every allocation has at least one stabilizing and one destabilizing anchor. At 1,058 equalized allocations of 493 baseline techniques, 5,290 random changes of the capital vector all moved the two non-anchored rates in opposite directions with the anchored rate between them, and for new capital placed in either non-anchored sector (2,116 cases) the criterion $(\rho_h - r^*)(\sigma_h - \sigma_l) > 0$ gave the sign of $d(r_h - r_l)$ without exception. For n = 4, 5 and 6 under the closure of Section S1.9.6, (S14) holds to $2 \times 10^{-9}$ at about 36,000 equalized allocations. Techniques with non-basic commodities are not covered by the proofs. Numerically, at 207,308 equalized allocations of such techniques with positive Sraffa prices, excluding the degenerate case in which the standard system consists of the anchored sector's commodity alone (every allocation then lies on the singular line of anchor a), the relative price moved in the direction of g′, and the receiver's price fell on at least one side of every stable allocation. At the 102,116 stable allocations among them, the receiver's rate never rose: it fell, except at 3,754 allocations where it remained unchanged to first order. In these cases the standard system reduces to a single commodity, produced from itself alone by one of the two sectors exchanging capital; the reference weights of the other sectors are then zero, and step (b) of the proof of Proposition 1 gives a zero derivative for that sector. Where some Sraffa prices are zero, some equalized prices vanish and the results can fail.

*Away from equalization.* For the paper's technique, every stabilizing first transfer (one that narrows $r_h - r_l$) on a grid of step 0.01 lowered both $p_h$ and $r_h$ (3,967 of 3,967), although the converse does not hold. For 2,962 randomly drawn viable techniques of the baseline family, the receiver's rate fell in all 13,365 stabilizing first transfers at random allocations. Its price fell in all but 47 (0.35 percent), in which its unit cost rose faster than its price, and its price relative to the donor's fell in 12,729 (95.2 percent). The fall of the receiver's rate is not universal, however. In a wider search of 148,464 stabilizing transfers across five families of techniques, with allocations near the edges of the simplex and near the standard proportions, it failed in 7 cases. All seven arise in nearly decomposable techniques, in which one sector is almost isolated from the other two, and far from equalization: $(r_h - r_l)/\bar{r}$ is 30 percent or more, and transformation coefficients range from 0.27 to 2.43; two of these cases have r* below 0.4, the other five between 1.9 and 2.8. The receiver's rate then rises slightly while the donor's rises 2 to 133 times faster. Formula (S11) shows why the results are local. Since $c^p \cdot Q = K_T$, $\sigma_h - \sigma_l = (q^*_h/Q_h - q^*_l/Q_l)K_T/(c^p \cdot q^*)$, so that the receiver's premium is

$$r_h - \bar{r} = \frac{\Gamma}{K_T Q_h \left( q_h^* / Q_h - q_l^* / Q_l \right)} \cdot \frac{c^p \cdot q^*}{c_h^p}:$$

a factor in quantities and values, proportional to Γ, times a ratio of unit costs. Near equalization Γ vanishes, so that the movement of that ratio does not matter; away from it, the movement is multiplied by the premium and can outweigh the first factor.

*Open questions.* Conditions guaranteeing a fall of the receiver's absolute price on both sides of z* or away from it, a lower bound on the size of the basin of attraction, other closures with more than three sectors and other selection rules remain open.

*S1.10 Falsifiable empirical implications*

The model suggests a falsifiable directional sequence for sectors observed over intervals in which technique and the real wage are sufficiently stable, or after controlling for their variation: above-average profitability should be followed by relative capital inflow, higher real output, compression of the sector's selling price relative to its input-price cost, and contraction of its profit-rate differential. The most robust object is $p_i/c_i^p = 1 + r_i$ rather than the absolute price alone. The analysis also predicts episodes that an excess-supply explanation cannot account for, since excess supply would lower the price of an expanding sector: a sector that receives capital and expands its output while its selling price rises and its profit rate falls, because its unit cost rises faster (Section S1.9). Such episodes should be rare, but observing them without accumulating inventories would favor the value-conservation account of the fall in the receiver's rate. The model's iterations are ordered adjustment states, not fixed calendar periods, so empirical work would have to estimate economically plausible lags.

More generally, industry data can confront the two accounts. Inventory-to-sales ratios and capacity utilization measure excess supply; input–output tables give the technique, hence the values, r* and the standard proportions from which the model predicts how a receiving industry's rate responds to the capital it receives (Sections S1.3 and S1.9). Regressing the change in that industry's profit-rate differential on both would show whether its fall tracks excess supply, as in classical gravitation analyses (Section S4.6), or the model's prediction. The model's states are recorded after realization and leave the transaction process between them unspecified, so a small, transient excess supply is not excluded: the test bears on whether excess supply accounts for the size of the fall, not on its strict absence. Demand shocks, which raise both investment and prices, must be controlled for, and the sample restricted to industries whose profit rates gravitate around a common value, which in the United States excludes the most capital-intensive industries (Duménil and Lévy 2002).

The aggregate equality of total profit and total surplus value holds by construction in the model and is therefore not itself a test. Its empirical content lies in the claim it makes about actual economies: that, net of fixed capital, foreign trade and unproductive sectors, aggregate profit should track aggregate surplus value even as relative prices depart substantially from values — a claim that data can contradict. The documented closeness of production prices to labor values (Shaikh 1984; Ochoa 1989; Cockshott and Cottrell 1997) is an indication, though not a proof, consistent with such a conservation constraint; a systematic, unexplained aggregate divergence would tell against it. The aggregate profit and exploitation rates, anchor-invariant at a given allocation and nearly so along the paths (Section 5.1 of the paper), are the robust objects; the relative-price structure, carrying the demand-side degree of freedom, is expected to be more variable.

Sections S1.8–S1.9 sharpen this implication. With input–output tables, labor inputs and the real-wage basket, r*, the standard proportions and the value-based rate $S/K_T$ can be measured. If the two aggregate equalities hold, identity (S10) implies a floor on the dispersion of sectoral profit rates, whatever the number of sectors and the closure: by the Cauchy–Schwarz inequality, the capital-weighted standard deviation of the rates is at least $\left|r^{*}-S/K_T\right|/\chi$ when $\chi>0$, where

$$\chi^2=\sum_j \frac{\left(\varphi_j^{*}-\varphi_j\right)^2}{\varphi_j}$$

measures the distance between actual and standard proportions.

Indeed, $\sum_j \left(\varphi_j^* - \varphi_j\right) = 0$ allows (S10) to be centered at $\bar{r}$, giving

$$\left|r^* - \bar{r}\right| = \left|\sum_j \left(\varphi_j^* - \varphi_j\right)\left(r_j - \bar{r}\right)\right| \leq \chi \left[\sum_j \varphi_j \left(r_j - \bar{r}\right)^2\right]^{1/2}.$$

If $\chi = 0$, the two averages coincide and this expression supplies no positive lower bound.

At $K_{init}$ the floor is 0.00327 (0.00326 under the iron anchor), against actual dispersions of 0.00328, 0.00439 and 0.00760 under the bread, iron and meat anchors: the bound is almost attained under the bread anchor. The mechanism further predicts that a reallocation of capital changes the aggregate profit rate according to the value rates $\rho_j$ of the sectors involved rather than their price rates, and that a sector's premium erodes faster when the capital it attracts brings $S/K_T$ closer to r*.

*Computational basis*

All numerical checks reported here use the synchronous reference implementation value_conservation_simulation.py and the Phase-0 calibration of the main paper. The derivatives of Section S1.4, the sequence test of Section S1.5 and the scan of Section S1.6 are produced by the companion script supplementary_verification.py; Sections S1.8–S1.10, Tables S5–S6 and Figure S1, and Figure 4 of the paper, are produced by anchor_analysis.py, generate_figure_S1.py and generate_figure4.py; Table S7 and the anticipation variant (Section S4.7) are produced by regulating_capital_rule.py and anticipation_rule.py. The numerical checks of Section S1.9 are produced by relative_price_theorem.py, which computes exact derivatives by complex step, except the counts of stabilizing first transfers, which are produced by anchor_analysis.py. These scripts import the same implementation and are archived with it on Zenodo (https://doi.org/10.5281/zenodo.22933924). The analytical equilibrium rate is r* = 0.2032510966. Numerical derivatives were computed by symmetric perturbation of the capital allocation while resolving the full price system at each perturbed state.

## S2 Value Conservation Tables Under Equilibrium and Technological Diffusion

### *S2.1 Pre-innovation equilibrium at iteration 199*

### Table S1 Sectoral aggregates in value terms

| | B | F | M | S | W |
|---|---|---|---|---|---|
| **BR1** | 367.935479 | 68.248989 | 123.815532 | 147.568229 | 707.568229 |
| **BR2** | 120.588953 | 83.880951 | 46.920511 | 28.212756 | 279.603171 |
| **BR3** | 94.834941 | 63.523487 | 30.251157 | 27.470111 | 216.079696 |
| **TOTAL** | **583.359373** | **215.653427** | **200.987200** | **203.251097** | **1203.251097** |

### Table S2 Corresponding price expressions and profits

| | $B^p$ | $F^p$ | $M^p$ | Π | $W^p$ |
|---|---|---|---|---|---|
| **BR1** | 346.012569 | 75.095502 | 131.901013 | 112.399703 | 665.408787 |
| **BR2** | 113.403833 | 92.295612 | 49.984544 | 51.968051 | 307.652039 |
| **BR3** | 89.184336 | 69.895953 | 32.226638 | 38.883343 | 230.190270 |
| **TOTAL** | **548.600738** | **237.287067** | **214.112194** | **203.251097** | **1203.251097** |

**Sectoral profit rates:** (0.2032511, 0.2032511, 0.2032511).

**Transformation coefficients:** (0.9404164, 1.1003167, 1.0653026).

**Transformation coefficients under meat anchoring** (K* = (561.9018, 258.0982, 180), iteration 199): (0.9404468, 1.1003522, 1.0653371). The two anchors therefore differ only in the fifth decimal.

Total price equals total value and total profit equals total surplus value to machine precision. The uniform rate matches the Perron–Frobenius rate r* = 0.20325110 to eight digits.

*S2.2 Mid-diffusion state at iteration 251*

Bread anchor, rising wages, θ = 0.5.

**Table S3 Sectoral aggregates in value terms**

| | **B** | **F** | **M** | **S** | **W** |
|---|---|---|---|---|---|
| **BR1** | 368.386851 | 67.287354 | 124.325795 | 148.436252 | 708.436252 |
| **BR2** | 120.161868 | 83.803554 | 46.542351 | 26.518524 | 277.026297 |
| **BR3** | 94.483292 | 62.319756 | 30.178660 | 27.495506 | 214.477214 |
| **TOTAL** | **583.032011** | **213.410664** | **201.046806** | **202.450283** | **1199.939763** |

**Table S4 Corresponding price expressions and profits**

| | $B^p$ | $F^p$ | $M^p$ | Π | $W^p$ |
|---|---|---|---|---|---|
| **BR1** | 345.618229 | 74.571755 | 132.322602 | 112.137853 | 664.650439 |
| **BR2** | 112.735109 | 92.875968 | 49.536020 | 51.869542 | 307.016638 |
| **BR3** | 88.643631 | 69.066374 | 32.119793 | 38.442887 | 228.272685 |
| **TOTAL** | **546.996969** | **236.514097** | **213.978415** | **202.450283** | **1199.939763** |

**Sectoral profit rates:** (0.20295982, 0.20329270, 0.20251240).

**Transformation coefficients:** (0.93819372, 1.10825810, 1.06432138).

Profit rates differ during diffusion, but the two aggregate equalities remain exact. Sectoral profits therefore differ from sectoral surplus values while their totals coincide.

## S3 Implementation of the Diffusion Mechanism

This section records the bookkeeping conventions used in the diffusion phase. The complete executable implementation is provided in value_conservation_simulation.py.

*S3.1 Allocation of Sector 2 capital between techniques*

$K_2\alpha = (1 - \theta)K_2$ and $K_2\beta = \theta K_2$

The number of units produced is $Q_2\alpha = K_2\alpha/k_2\alpha$ and $Q_2\beta = K_2\beta/k_2\beta$, with $Q_2 = Q_2\alpha + Q_2\beta$. Thus θ is a share of Sector 2 capital, not a share of firms or output. Since the new technique saves capital ($k_{2\beta} < k_{2\alpha}$), a given fraction of capital yields more units under it, so the share of physical output produced with it slightly exceeds θ (by at most 0.22 percentage points, at θ = 0.5); θ is nonetheless the weight used to interpolate the per-unit coefficients, so $\Lambda_{calc}$ is capital-weighted rather than the realized social average, the two differing only during diffusion and by at most 0.22 points.

*S3.2 Aggregate exploitation rate of Sector 2*

$e_2 = S_2/V_2$

$S_2 = Q_2\alpha s_2\alpha + Q_2\beta s_2\beta$ and $V_2 = Q_2\alpha\ wage_2\alpha + Q_2\beta\ wage_2\beta$. Both techniques produce iron commanding the same socially recognized unit value, but their costs and surplus values differ.

*S3.3 Revaluation of Sector 2 capital*

$K_2(t) = K_2(t-1)\{[1-\theta(t)]k_2\alpha(t)/k_2\alpha(t-1) + \theta(t)k_2\beta(t)/k_2\beta(t-1)\}$

Revaluation is applied at the beginning of iteration t, before aggregates, prices, profit rates and intersectoral migration are computed. Every recorded row is therefore synchronous. The weights use θ(t) rather than θ(t − 1), as in the code, so that the revalued $K_2$ is not exactly the sum of the separately revalued capitals committed to each technique in the previous state.

*S3.4 Real wage schedule*

v(t) = v(t−1) + v_add for t in [202 + delay, 301].

The optional delay postpones the first wage increment. The paper's rising-wage scenarios use v3_add = 0.000354 and delay = 0, giving 100 increments and cumulative $\Delta v_3 = 0.0354$.

## S4 Additional Theoretical and Numerical Details

*S4.1 Simultaneous valuation and realization*

Our axiomatic use of value conservation differs methodologically from the Temporal Single System Interpretation (TSSI; Kliman 2007). We employ simultaneous determination, $\Lambda = \ell(I-A)^{-1}$, to describe the value system at each recorded state. This is not temporal valuation of inputs at historical purchase prices: circulating capital is valued using the operative labor values, and Sector 2 committed capital is revalued during diffusion as specified in Section S3.3. Unlike the TSSI's single-system framework, values are determined here by the technique, independently of prices, and the two aggregate equalities link the two valuations at every state; the simultaneous determination also supplies what the TSSI's rejection of equilibrium withholds, a coherent equilibrium concept compatible with value conservation. The circulating-capital case treated here isolates that equilibrium basis in its purest form; fixed capital, which introduces the temporal input/output differentiation the TSSI emphasizes, is treated in Ankri (forthcoming).

The tabulated magnitudes assume successful realization. Marx's salto mortale concerns validation of commodities through sale. A recorded state represents the outcome after supply and demand have met, while capital allocation and sectoral profitability need not yet be at their uniform-profit equilibrium. Inventories, shortages, bargaining and losses during the intervening market adjustment are outside the sequence. If commodities cannot be sold on the assumed terms, that state is not described by the full-realization accounts. Physical reproducibility and the two invariants are necessary restrictions under the model's hypothesis; without specified demand they do not ensure sales. Iterations order selected realized states; they do not measure calendar periods.

**Conservation law versus the New Interpretation.** Our framework treats Marx's two aggregate equalities as a conservation law on gross aggregates: transformation redistributes surplus value between sectors — sectoral profit departs from sectoral surplus value — while conserving both aggregate totals: total price equals total value, and total profit equals total surplus value. This differs from the New Interpretation (Duménil 1980; Foley 1982), which preserves the equivalence between living labor and monetary value added — a net-product

condition — rather than the gross totals. The two do not coincide: for the pre-innovation bread-anchor equilibrium, total value added is 384.1 in value terms but 381.4 in price terms (a 0.7 percent gap), whereas the gross equality (total price = total value) holds exactly. What our law conserves is therefore the gross total Marx himself states, inclusive of transferred value, not the net product. The small size of the value-added gap indicates that the dynamic results reported here are not artifacts of this choice. Treating the two equalities as a conservation law, rather than as a numéraire-fixing normalization, is what distinguishes the present construction from the New Interpretation, which conserves the labor–value-added equivalence rather than Marx's gross aggregate equalities.

**Conservation within states, devaluation between them.** Duménil and Lévy (2000) reject the value conservation principle of Freeman's sequential approach, in which inputs transfer their historical labor content to outputs and value is conserved from period to period: individual commodities may lose value, but each loss is compensated by a corresponding gain for another commodity. Against it, they argue that technical progress devalues capital economy-wide, because labor embodied in the past that is no longer socially necessary vanishes, whereas circulation redistributes value without creating or destroying it. The conservation imposed here is of the latter kind. It holds within each recorded state, between the value and price accounts of the same commodities, and governs the redistribution of surplus value through prices; it is not a conservation of value across states. The "strict conservation of value" of our title refers to this within-state law, not to an intertemporal principle. Because values are simultaneous, technical change revalues committed capital: during diffusion, total capital falls by about 0.5 percent while output rises (Sections 3.2 and 4.1 of the paper). The model is therefore consistent with their account of economy-wide devaluation. Its difference from the New Interpretation, discussed above, lies in which aggregates are conserved within each state.

*S4.2 Composition of the net product and accumulation ceilings*

Let $D_i$ denote the value of the physical commodity i required for replacement and the wage bundle, and $N_i = W_i - D_i$ the remaining net product. At fixed technique and real wages, a proportional expansion by g requires $gD_i$ of each commodity. Without trade, substitution or prior stocks, feasibility requires $g D_i \leq N_i$ for each commodity, hence $g \leq min_i(N_i / D_i)$. This is a ceiling on a hypothetical balanced expansion, not an accumulation process simulated across iterations.

Under the bread anchor at iteration 199, D = (583.36, 215.65, 200.99) and N = (124.21, 63.95, 15.09). The ratios $N_i/D_i$ are (0.2129, 0.2965, 0.0751), so the maximum uniform expansion is 7.51 percent, limited by meat. It absorbs $g\Sigma_i D_i/S$ = 36.95 percent of total surplus value.

Under the meat anchor at iteration 199, D = (583.50, 215.22, 201.28) and N = (126.47, 71.84, 4.94). The ratios $N_i/D_i$ are (0.2168, 0.3338, 0.0245), so the maximum uniform expansion is 2.45 percent, limited by meat. It absorbs $g\Sigma_i D_i/S$ = 12.07 percent of total surplus value.

Thus the rounded ceilings are 7.5 percent and 2.5 percent, absorbing approximately 37 percent and 12 percent of surplus value. Accumulation alone cannot realize the whole net product under these proportional-expansion assumptions. Consumption or exchange is also required. Equivalent external exchange can preserve the domestic value accounts; transfers through unequal exchange require a wider accounting boundary. The composition of the net product, not merely its total value, constrains feasible accumulation. Simple reproduction imposes different restrictions; balanced proportions do not arise automatically from the present adjustment rule.

*S4.3 Automation and the source of surplus value*

For a homogeneous technique evaluated at its corresponding labor values, let Λv be the value of the real wage per labor hour. Since $\Lambda = \Lambda A + l$ and $k = \Lambda A + (\Lambda v) l$, subtraction gives $s = \Lambda - k = (1 - \Lambda v) l$, or $s_j = (1 - \Lambda v) l_j$. Living labor contributes new value; material inputs transfer their value. A machine-maintenance bundle therefore enters the technical matrix A rather than the wage vector v, even if its physical composition resembles workers' consumption. This is an axiom of the value accounting, not a conclusion proved by simulation. During coexistence of techniques, individual surpluses are instead computed as $s_2\alpha = \Lambda_2 - k_2\alpha$ and $s_2\beta = \Lambda_2 - k_2\beta$ at the common social value.

With bread anchoring and constant real wages, Sector 2 surplus value falls from 28.21 at iteration 199 to 24.87 at iteration 451, while its profit changes from 51.97 to 52.02. Profit can exceed the branch's own surplus value because transformation redistributes surplus value across sectors. In the limiting thought experiment of full automation, a branch could receive profit without creating surplus value, provided other branches continue to generate it. Full automation itself is not one of the four simulated scenarios.

*S4.4 Off-equilibrium price constraints and Phase-0 numerical detail*

In our circulating-capital specification, let Q be gross physical output, $d = \tilde{A}Q$ the physical bundle used as productive inputs and workers' consumption, and $N = Q - d$ the physical surplus. For any relative-price vector $\hat{p}$, write $p = \gamma\, \hat{p}$. Normalizing total price to total value W fixes $\gamma = W/(\hat{p}Q)$, so $\Pi = \gamma\, \hat{p}N = W(\hat{p}N)/(\hat{p}Q)$. Hence $\Pi = S$ if and only if $(\hat{p}N)/(\hat{p}Q) = S/W$. Because γ cancels, this condition is numéraire-independent and restricts relative prices themselves. A normalization can impose either aggregate equality, but not in general both; Marx's two aggregate equalities therefore restrict admissible off-equilibrium price vectors rather than merely choosing a numéraire. Sraffa does not predict a particular off-equilibrium price path; the point is that the static physical system alone supplies no corresponding pair of aggregate invariants.

In Phase 0, with technique and real wage fixed, the Sraffian equilibrium rate remains 0.2032511, while the contemporaneous average rises from 0.2029634 to 0.2032511 as capital is reallocated. During diffusion two techniques coexist, no single augmented matrix describes the economy, and $\bar{r}$ differs by up to $2.1 \times 10^{-5}$ from the rate implied by the interpolated technique. Along the path, $\bar{r}(t) = \Sigma s_j Q_j / \Sigma k_j Q_j$ closes the price system as a value magnitude.

*S4.5 Real-wage composition and profitability (Duménil 1982)*

In a Morishima-type system the composition of the real-wage basket, not only its value, affects profitability. Holding $\Lambda v$ and hence exploitation fixed at 1.1235674, a bread-only basket gives $r^* = 0.2135$ and a meat-only basket 0.1873, against 0.2033 for the reference basket. Consumption norms therefore shape the material structure of reproduction and profitability, which is why the real wage enters as a physical vector rather than a scalar share.

*S4.6 Quasi-static states and classical price gravitation*

Classical gravitation analyses describe capital mobility together with market adjustment: above-average profitability attracts capital, and changes in supply, demand and inventories affect prices (Duménil and Lévy 1987). The present construction records selected states after market realization. Along its convergent trajectories for the present technique, capital inflow expands output while the receiver’s price and profitability premium fall. Sharing this direction of movement does not identify the transaction process through which the calculated prices emerge.

The two aggregate equalities constrain prices and quantities jointly. At a given allocation they leave one degree of freedom, so conservation does not determine a unique realized price vector. Setting one sector's rate at the contemporaneous average selects a particular trajectory through the admissible set. It makes no claim about that sector's intention: a firm may seek a higher margin or lower its price to gain market share. Such strategies and demand conditions may select different outcomes. Under the conservation hypothesis, any resulting realized state must satisfy the aggregate equalities and physical reproduction conditions; those restrictions alone do not guarantee its realization. Whatever the selection, a uniform rate of profit can be reached only on the line $S = r^*K_T$, at Sraffa's prices (Section S1.8).

Capital mobility changes the conditions for the next compatible state. At fixed technique it changes output proportions and the amount of living labor employed, hence total surplus value even when total capital is constant. The two equalities then require a corresponding redistribution through prices and profits. Section S1.8 measures the discrepancy from uniform profitability by Γ and identifies the excess profits relative to the current average. At $K_{init}$, under the bread anchor, the iron-to-meat transfer lowers the social organic composition toward that of the standard system and narrows Γ sufficiently to reduce the receiver's premium. Its profit grows more slowly than output, reducing profit per unit and the selling price. Elsewhere the change in relative proportions can dominate. The response thus depends on the allocation and selection rule, not only on the receiving sector's greater output.

Marx distinguishes supply-and-demand fluctuations from the determination of value and prices of production, while describing profit-rate equalization through adjustments of supply to social need (Marx [1894] 1981, ch. 10). Here market realization at each state precedes any subsequent equalization of profitability. The final center of gravitation, the uniform-profit price vector, is reached only on a convergent trajectory. Before that point, realized prices can differ from it even though the recorded quantities have sold. Because every recorded state is assumed realized, demand is taken to have absorbed each of these price-quantity pairs; the pairs nevertheless trace the selected compatible sequence, not a demand curve, and the model does not establish whether buyers adjust quantities to value-determined prices or prices respond to demand.

Stability adds a restriction on the selected sequence. Near equalization, for any viable technique in which every commodity is basic, stability is equivalent to a fall in the receiver's price relative to the donor's, and a stable approach lowers the receiver's profit rate, even in the rare cases where its absolute price rises (Section S1.9). The direction of movement is therefore that of classical gravitation, but the explanation differs. No excess-supply price rule is involved: near equalization, the aggregate equalities and the anchor fix the change in profit rates, which reflects the narrowing of the gap between surplus value produced and the profit a uniform rate requires, and the change in prices follows. The fall in the receiver's price, when it occurs, is a consequence of this redistribution rather than its cause. Steedman (1984) showed that a sector whose market price exceeds its normal price may nevertheless earn a below-average rate of profit, if the prices of its inputs are proportionally higher still; Bellino and Serrano (2018) observe that this cannot hold in every sector at once. Section S1.9 finds a related case near equalization: on one side of the equalized allocation, the receiver's price rises while its unit cost rises faster, and its rate still falls. For the present technique, every locally stabilizing first transfer also lowers the receiver's price and every price increase accompanies destabilizing feedback, although a price fall alone does not ensure stability; for other techniques, the absolute-price implication can fail on one side of the equalized allocation or away from it (Section S1.9). A divergent trajectory cannot accomplish profit-rate equalization under the stipulated rules and can leave the admissible domain. This is a property of the quasi-static selection, not a claim that every rise in an actual sector's output and price must be unsustainable. The analysis separates restrictions on realized states from the market processes that select them, while leaving those processes unmodelled.

*S4.7 Alternative migration rules: the best technique (regulating capital) and anticipation*

Shaikh (2016, ch. 7) argues that competition equalizes the profit rates of regulating capitals, the best generally reproducible conditions of production, which attract new investment, rather than branch averages. During diffusion, Sector 2 operates two techniques (Section S3.1). In this variant, only the migration rule changes: it represents Sector 2 by the profit rate of its new technique at the prices of the current state:

$$r_{2\beta}=\frac{x_2\Lambda_2-c_{2\beta}}{c_{2\beta}}$$

$$c_{2\beta}=x_1 b_{2\beta}+x_2 f_{2\beta}+x_3 m_{2\beta}$$

where $b_{2\beta}$, $f_{2\beta}$ and $m_{2\beta}$ are the values of the bread (wage bread included), iron and meat (wage meat included) used per unit of iron by the new technique. The rate $r_{2\alpha}$ of the old technique is defined likewise, and the two rates average to $r_2$ with weights equal to each technique's costs. Transformation, anchor, revaluation and wage path are unchanged, and the conservation residuals stay below $3 \times 10^{-13}$. Table S7 compares the two rules; the script regulating_capital_rule.py reproduces it.

**Table S7 Migration on the branch average (the paper's rule) and on the rate of the best technique**

| Anchor | Flow factor | Migration signal for Sector 2 | Final $\bar{r}$ | Final $K_2$ | Lowest $r_2 - \bar{r}$ | Highest $r_{2\beta} - \bar{r}$ after ten iterations | Lowest $r_{2\alpha} - \bar{r}$ |
|---|---|---|---|---|---|---|---|
| Bread | 10 | Branch average | 0.20551 | 248.1 | 0.00 | +0.85 | −0.93 |
| Bread | 10 | Best technique | Diverges | — | — | — | — |
| Bread | 5 | Branch average | 0.20551 | 248.1 | 0.00 | +0.85 | −0.90 |
| Bread | 5 | Best technique | 0.20551 | 237.7 | −0.85 | +0.38 | −1.04 |
| Meat | 10 | Branch average | 0.20551 | 252.6 | 0.00 | +0.84 | −0.94 |

| Anchor | Flow factor | Migration signal for Sector 2 | Final $\bar{r}$ | Final $K_2$ | Lowest $r_2 - \bar{r}$ | Highest $r_{2\beta} - \bar{r}$ after ten iterations | Lowest $r_{2\alpha} - \bar{r}$ |
|---|---|---|---|---|---|---|---|
| Meat | 10 | Best technique | 0.20551 | 258.5 | −0.58 | +0.33 | −0.96 |
| Meat | 5 | Branch average | 0.20551 | 252.6 | −0.01 | +0.84 | −0.94 |
| Meat | 5 | Best technique | 0.20551 | 255.9 | −0.53 | +0.46 | −0.97 |

*Note:* Constant real wage. Rate differences in percentage points, during diffusion (iterations 202–301); "after ten iterations" covers iterations 212–301. Sector 2 holds 251.4 (bread anchor) or 258.1 (meat anchor) when diffusion starts.

Three results hold in every convergent run. First, the final uniform rate is that of the paper: 0.20551 at a constant real wage and, with the rising-wage path, 0.20267. Second, during diffusion Sector 2's average rate falls below $\bar{r}$, by up to 0.85 percentage points, whereas it stays within 0.01 point of $\bar{r}$ under the paper's rule; the old technique earns about one point less than $\bar{r}$; and the new technique's premium, about 0.95 points at the onset, falls within ten iterations to 0.3–0.5 points and vanishes as diffusion completes. Competition thus nearly equalizes the rate of the best technique while the branch average, weighed down by older capital, remains below it: the rule reproduces the pattern Shaikh describes, in which regulating rates equalize and branch averages need not. Third, the final allocation differs from the paper's: equalized allocations form a one-parameter family (Section 5.1 of the paper), and the path selects the point.

The movement of capital depends on the anchor. Under meat anchoring, Sector 2 attracts capital during diffusion: $K_2$ rises from 258.1 to 260.5 and ends at 258.5, whereas it declines steadily to 252.6 under the paper's rule. Under bread anchoring with the paper's flow factor (10), the first inflows raise the meat sector's rate so strongly that capital then leaves Sector 2 cumulatively, and $K_2$ vanishes near iteration 334. With a flow factor of 5 the path converges; Sector 2 gains capital only briefly ($K_2$ peaks at 252.4) and ends with 237.7, against 248.1 under the paper's rule. The same pattern holds with the rising-wage path.

A second variant tests anticipation. Capital moves on anticipated rates $r(t) + \tau[r(t) - r(t - 1)]$, where $\tau$ is an anticipation horizon in iterations, so that migration responds to the change in each rate as well as to its level; equalized states are unchanged, since anticipated and actual rates then coincide. Under bread anchoring, for $\tau = 1$, 2 and 5, the final allocation and uniform rate are those of the paper under both wage regimes, but the Phase-0 dispersion falls below $10^{-6}$ only at iterations 142, 153 and 185, against 131; for $\tau = 10$, rates no longer equalize. Under meat anchoring, $\tau = 1$ converges in Phase 0 (iteration 31, against 18) but rates do not re-equalize after the innovation, and for $\tau \geq 2$ a profit rate becomes non-positive during Phase 0. In every run that converges, Sector 2's average rate stays within 0.03 point of $\bar{r}$ during diffusion. Anticipation thus slows equalization and, when strong, destabilizes it; it does not produce the pattern of Table S7, which follows from the signal chosen, not from its timing. The script anticipation_rule.py reproduces these results.